\documentclass[preprint]{aastex}

\begin{document}

\voffset -0.5in

\shortauthors{Welty, Hobbs, \& Morton}
\shorttitle{Interstellar \ion{Ca}{1}}

\slugcomment{accepted to ApJS}  

\title{High-Resolution Observations of Interstellar \ion{Ca}{1} Absorption ---
Implications for Depletions and Electron Densities in Diffuse Clouds}

\author{Daniel E. Welty\altaffilmark{1,2,3}}
\affil{University of Chicago, Astronomy and Astrophysics Center, 5640 S. Ellis Ave., Chicago, IL  60637}
\email{welty@oddjob.uchicago.edu}
\altaffiltext{1}{Visiting observer, Kitt Peak National Observatory}
\altaffiltext{2}{Visiting observer, McDonald Observatory}
\altaffiltext{3}{Visiting observer, Anglo-Australian Observatory}

\author{L. M. Hobbs\altaffilmark{2}}
\affil{University of Chicago, Yerkes Observatory, Williams Bay, WI  53191-0258}
\email{hobbs@yerkes.uchicago.edu}

\author{Donald C. Morton\altaffilmark{3}}
\affil{Herzberg Institute of Astrophysics, National Research Council, 5071 W. Saanich Rd., Victoria, BC, Canada V9E 2E7}
\email{don.morton@nrc.ca}

\begin{abstract}

We present high-resolution (FWHM $\sim$ 0.3--1.5 km~s$^{-1}$) spectra, obtained with the AAT UHRF, the McDonald Observatory 2.7-m coud\'{e} spectrograph, and/or the KPNO coud\'{e} feed, of interstellar \ion{Ca}{1} absorption toward 30 Galactic stars.
Comparisons of the column densities of \ion{Ca}{1}, \ion{Ca}{2}, \ion{K}{1}, and other species --- for individual components identified in the line profiles and also when integrated over entire lines of sight --- yield information on relative electron densities and depletions (dependent on assumptions regarding the ionization equilibrium).
There is no obvious relationship between the ratio $N$(\ion{Ca}{1})/$N$(\ion{Ca}{2}) [equal to $n_e$/($\Gamma$/$\alpha_r$) for photoionization equilibrium] and the fraction of hydrogen in molecular form $f$(H$_2$) (often taken to be indicative of the local density $n_{\rm H}$).
For a smaller sample of sightlines for which the thermal pressure ($n_{\rm H}T$) and local density can be estimated via analysis of the \ion{C}{1} fine-structure excitation, the average electron density inferred from C, Na, and K (assuming photoionization equilibrium) seems to be independent of $n_{\rm H}$ and $n_{\rm H}T$.
While the electron density ($n_e$) obtained from the ratio $N$(\ion{Ca}{1})/$N$(\ion{Ca}{2}) is often significantly higher than the values derived from other elements,
the patterns of relative $n_e$ derived from different elements show both similarities and differences for different lines of sight --- suggesting that additional processes besides photoionization and radiative recombination commonly and significantly affect the ionization balance of heavy elements in diffuse interstellar clouds.
Such additional processes may also contribute to the (apparently) larger than expected fractional ionizations ($n_e$/$n_{\rm H}$) found for some lines of sight with independent determinations of $n_{\rm H}$.
In general, inclusion of ``grain-assisted'' recombination does reduce the inferred $n_e$, but it does not reconcile the $n_e$ estimated from different elements; it may, however, suggest some dependence of $n_e$ on $n_{\rm H}$.
The depletion of calcium may have a much weaker dependence on density than was suggested by earlier comparisons with CH and CN.
Two appendices present similar high-resolution spectra of \ion{Fe}{1} for a few stars and give a compilation of column density data for \ion{Ca}{1}, \ion{Ca}{2}, \ion{Fe}{1}, and \ion{S}{1}.

\end{abstract}

\keywords{ISM: abundances, ISM: atoms, ISM: kinematics and dynamics, Line: profiles}

\section{Introduction}
\label{sec-intro}

High-resolution (FWHM $\sim$ 0.3--1.8 km~s$^{-1}$) surveys of the optical interstellar absorption lines of \ion{Na}{1} (Welty, Hobbs, \& Kulkarni 1994; hereafter WHK), \ion{Ca}{2} (Welty, Morton, \& Hobbs 1996; hereafter WMH), and \ion{K}{1} (Welty \& Hobbs 2001; hereafter WH) have revealed complex velocity structure in many Galactic lines of sight.
Fits to the absorption-line profiles have yielded column densities, line widths, and velocities for several hundred individual components in each case --- enabling statistical studies of the properties of the corresponding individual interstellar clouds.
Under the usual assumption of photoionization equilibrium, comparisons of such high-resolution spectra of \ion{K}{1} and \ion{Ca}{2} with similar spectra of \ion{Ca}{1} can provide information on the relative electron densities and elemental depletions in the individual components.
This information can then be used to model the profiles of many other neutral and singly ionized species seen in lower resolution UV spectra --- yielding detailed abundances, depletions, and physical conditions for the individual clouds (e.g., Welty et al. 1999b).

Because even the strongest line of \ion{Ca}{1} (at 4226.7\AA) is typically quite weak (W$_{\lambda}$ $<$ 10 m\AA), however, very few spectra at resolutions better than 3 km~s$^{-1}$ have ever been reported (Hobbs 1971; Welty et al. 1999b).
The earliest quantitative comparisons of \ion{Ca}{1} and \ion{Ca}{2} yielded mean electron densities ($n_e$) ranging from about 0.1--1.3 cm$^{-3}$ (Hobbs 1971; White 1973; Federman \& Hobbs 1983; but using the slightly different atomic data adopted in this study).
If a fractional ionization $n_e$/$n_{\rm H}$ comparable to the interstellar carbon abundance is assumed (i.e., if photoionization of \ion{C}{1} is the primary source of electrons), such values of $n_e$ would imply fairly high local hydrogen densities ($n_{\rm H}$) and thermal pressures ($n_{\rm H}T$).
Alternatively, if the true $n_{\rm H}$ are lower, then those $n_e$ might suggest that hydrogen is slightly ionized even in predominantly neutral clouds.

While ratios such as $N$(\ion{Ca}{1})/$N$(\ion{Ca}{2}) can in principle yield estimates for $n_e$, recent observational and theoretical studies have suggested that additional processes besides photoionization and radiative recombination may commonly and significantly affect the ionization balance of heavy elements in diffuse clouds.
For example, the electron densities derived from different $N$(\ion{X}{1})/$N$(\ion{X}{2}) ratios in a given line of sight can differ by factors of 10 or more (Fitzpatrick \& Spitzer 1997; Welty et al. 1999a, 1999b).
Moreover, the corresponding fractional ionizations $n_e$/$n_{\rm H}$ (with $n_{\rm H}$ estimated from the \ion{C}{1} fine-structure excitation equilibrium) can be significantly higher than the value ``expected'' from photoionization of carbon.
While some additional electrons may arise from cosmic ray and/or x-ray ionization of hydrogen (e.g., Wolfire et al. 1995), it has been suspected for some time that electron densities derived under the assumption of photoionization equilibrium may be spuriously high (e.g., Snow 1975).
Another potentially relevant process is charge exchange between the various dominant singly ionized species and neutral or negatively charged small grains (or large molecules), which would enhance the corresponding trace neutral species --- thus yielding spuriously high $n_e$ if only radiative recombination is assumed (Lepp et al. 1988; Weingartner \& Draine 2001; Liszt 2003).
Accurate abundances of both neutral and singly ionized species --- for a number of elements and for sightlines sampling a variety of interstellar environments --- are needed to identify and disentangle the various processes that may affect the ionization balance.

In this paper, we discuss spectra of interstellar \ion{Ca}{1}, observed at resolutions of 0.3--1.5 km~s$^{-1}$, toward 30 relatively bright Galactic stars.
The high spectral resolution enables more detailed, component-by-component comparisons with similar spectra of \ion{Na}{1}, \ion{K}{1}, \ion{Ca}{2}, and various molecular species --- and thus more accurate characterization of the properties of individual interstellar clouds.
In Sections~\ref{sec-obs} and~\ref{sec-res}, we describe the procedures used to obtain, reduce, and analyze the spectra.
In Section~\ref{sec-disc}, we discuss the statistical properties of the ensemble of individual components detected in \ion{Ca}{1}, we compare the \ion{Ca}{1} absorption with that due to other species, and we explore possible implications for determinations of the electron density and depletions in interstellar clouds.
Comparisons of various $N$(\ion{X}{1})/$N$(\ion{X}{2}) ratios for five lines of sight reveal both similarities and differences in the patterns of relative $n_e$ inferred from the different elements --- suggesting that several different processes may affect the ionization balance.
In Section~\ref{sec-sum}, we give a brief summary of the results.
In two appendices, we show similar high-resolution spectra of \ion{Fe}{1} toward six stars and we list available column density measurements for \ion{Ca}{1}, \ion{Ca}{2}, \ion{S}{1}, and \ion{Fe}{1} --- supplementing the values for H$_{\rm tot}$ (= \ion{H}{1} + 2 H$_2$), H$_2$, \ion{Na}{1}, \ion{K}{1}, \ion{Li}{1}, \ion{C}{1}, and CH tabulated by WH.

\section{Observations and Data Reduction}
\label{sec-obs}

\subsection{Program Stars}
\label{sec-stars}

The 30 stars observed in the present high-resolution \ion{Ca}{1} survey are listed in Table~\ref{tab:stel}.
All but one of the stars ($\alpha$ Cam) were included in at least one of our previous surveys (WHK, WMH, WH), where we have tabulated Galactic coordinates, $V$, $E(B-V)$, spectral types, and distances.
Because the \ion{Ca}{1} absorption is typically quite weak, most of the stars chosen for observation in this survey have $E(B-V)$ $>$ 0.05. 

\subsection{Instrumental Setups}
\label{sec-inst}

Because the instrumental setups employed for these observations of \ion{Ca}{1} are very similar to those used for the \ion{Na}{1}, \ion{Ca}{2}, and \ion{K}{1} surveys, we will give only brief summaries here; more extensive descriptions may be found in WHK and WMH.

High-resolution \ion{Ca}{1} spectra, with $\Delta v$ = 0.53$\pm$0.05 km~s$^{-1}$ (FWHM), were obtained for 10 stars with the McDonald Observatory (McD) 2.7-m telescope and coud\'{e} spectrograph in 1996 September.
As for the \ion{Na}{1} and \ion{K}{1} spectra reported by WHK and WH, the echelle spectrograph was used in the double-pass configuration. 
A slit width of 100$\mu$m, corresponding to 0.23 arcsec on the sky, was used for all observations.
The TK-4 CCD (1024 $\times$ 1024, 24 \micron~pixels) was placed at the ``scanner'' focus.
The dispersion achieved with this setup corresponds to 3.1 m\AA\ per pixel, or 2.4 pixels per resolution element.
A quartz lamp was used for flat-field exposures; a Th-Ar hollow cathode lamp was used to determine the wavelength scale.
The resolution was determined from the measured widths of the thorium lines in the Th-Ar exposures, assuming an intrinsic (thermally broadened) width of 0.55 km~s$^{-1}$ (Willmarth 1994, priv. comm.; Crawford et al. 1994; WHK).
A signal-to-noise ratio (S/N) $\sim$ 165 (per half resolution element) was achieved in 120 minutes (two 60-minute exposures) for $\xi$ Per, which has $B$ $\sim$ 4.1. 

Higher resolution \ion{Ca}{1} spectra, with $\Delta v$ = 0.32 km~s$^{-1}$ (FWHM), were obtained for 5 stars with the Ultra-High Resolution Facility (UHRF; Diego et al. 1995) on the 3.9-m Anglo-Australian Telescope in 1997 March.
The 35-slice image slicer, which takes light from a 1.0 $\times$ 1.5 arcsec aperture to form a 30$\mu$m wide pseudo-slit (Diego 1993), and the Tektronix CCD (1024 $\times$ 1024, 24 \micron~pixels) were used for all the observations reported here.
%All exposures were binned by 4 pixels perpendicular to the dispersion.
The dispersion achieved with this setup corresponds to 2.2 m\AA\ per pixel, or 2 pixels per resolution element.
Because of the large number of CCD rows filled by the stellar spectrum and the clean response of the CCD, flat fielding was not required (for the S/Ns achieved in this study).
The spectral resolution, $\Delta v$ = 0.32 km~s$^{-1}$, was determined via scans of a stabilized He-Ne laser line at 6328.160 \AA~ and by examination of the widths of the thorium lines in the Th-Ar exposures used for wavelength calibration.
Individual exposures were limited to 30 minutes to preserve the high spectral resolution; S/N $\sim$ 225 was achieved in three exposures totalling 75 minutes for $\delta$ Sco, which has B $\sim$ 2.2.

\ion{Ca}{1} spectra with a resolution $\Delta v$ = 1.3--1.5 km~s$^{-1}$ (FWHM) were obtained for 22 stars with the 0.9-m coud\'{e} feed (CF) telescope and coud\'{e} spectrograph of the Kitt Peak National Observatory during various runs from 1993 to 2000.
To achieve the desired high resolution, we used the echelle grating, camera 6, and a 90--120 \micron\ entrance slit, corresponding to 0.6--0.8 arcsec on the sky.
Several different grism, wedge, and filter combinations were used in the various runs to cross-disperse the spectra and to isolate the desired spectral orders. 
Two different CCD's were used (T1KA: 1024 $\times$ 1024, 24 \micron~pixels in 1993; F3KB: 1024 $\times$ 3072, 15 \micron~pixels thereafter).
The dispersion achieved with this setup corresponds to 8.3 m\AA\ per pixel in 1993 and 5.2 m\AA\ per pixel thereafter, corresponding to about 2.4 and 3.8 pixels per resolution element, respectively; the post-1993 spectra were re-binned to approximately optimal sampling.
Flat-field exposures were obtained using a quartz lamp; the wavelength calibration was established via a Th-Ar lamp.
We achieved S/N $\sim$ 220 in 100 minutes for $\xi$ Per.

\subsection{Data Reduction}
\label{sec-datred}

The initial processing of the CCD images from all three telescopes followed procedures similar to those used in our previous surveys (WHK, WMH, WH).
Various routines within IRAF were employed to subtract the bias and divide by a normalized flat field (except for the UHRF data); the one-dimensional spectra then were extracted using the apextract package, with variance weighting.
Wavelength calibration was accomplished via exposures of Th-Ar hollow cathode lamps, using the thorium rest wavelengths of Palmer \& Engelman (1983).
Multiple spectra for a given target were sinc interpolated to a common heliocentric wavelength grid and summed.

Comparisons between the individual Th-Ar exposures and between the \ion{Ca}{1} spectra and previously obtained high-resolution spectra of \ion{Na}{1}, \ion{K}{1}, and \ion{Ca}{2} yield complementary information on the accuracy of the velocities.
For the coud\'{e} feed spectra, quadratic fits to the positions of the 13--28 lines found in the \ion{Ca}{1} order yield typical residuals of 0.8 m\AA, or 0.06 km~s$^{-1}$.
Calibration exposures were usually obtained at the beginning, middle, and end of each night; in general, the zero points agree to within 5 m\AA, or 0.35 km~s$^{-1}$ (and usually much better), and the dispersions agree to within 0.02\% for any individual night.
At the higher McDonald and UHRF resolutions, only 4--5 lines with previously known wavelengths were generally found within the $\sim$ 1.5--2.3 \AA\ spectral range recorded on the CCDs. 
Linear fits to the positions of those lines yield typical residuals of 0.3--0.8 m\AA, or 0.02--0.06 km~s$^{-1}$.
For the UHRF observations, calibration spectra were obtained several times during each night. 
The zero points agree to within 5 m\AA, or 0.35 km~s$^{-1}$, and the dispersions agree to better than 0.05\% for each night.
Comparisons of the interstellar line profiles for seven lines of sight observed at both McDonald and the coud\'{e} feed suggest that any systematic differences in velocity zero point are less than 0.2 km~s$^{-1}$. 
There also appears to be good agreement in both velocity scale and zero point between these \ion{Ca}{1} spectra and the \ion{Na}{1}, \ion{Ca}{2}, and \ion{K}{1} spectra reported by WHK, WMH, and WH.
For example, for 35 ``corresponding'' components identified in both \ion{Ca}{1} and \ion{K}{1} in the profile analysis (see below), $v$(\ion{Ca}{1}) $-$ $v$(\ion{K}{1}) is 0.02 $\pm$ 0.19 km~s$^{-1}$ (mean $\pm$ standard deviation).

The summed, wavelength-calibrated spectra were normalized by fitting Legendre polynomials to regions free of interstellar absorption.
In all cases, sufficient continuum was present on both sides of the interstellar features.
The S/Ns determined in the continuum fits and the total equivalent widths of the interstellar \ion{Ca}{1} absorption measured from the normalized spectra are given in columns 2--3 (CF), 4--5 (McD), and 6--7 (UHRF) of Table~\ref{tab:stel}.
The median S/Ns achieved for the three data sets are $\sim$ 120, 175, and 215, respectively.
Those median S/Ns yield 2$\sigma$ equivalent width uncertainties or limits for narrow, unresolved absorption lines (including contributions from both photon noise and continuum placement uncertainty) of about 0.95 m\AA\ (CF), 0.3 m\AA\ (McD), and 0.16 m\AA\ (UHRF), which correspond to unsaturated \ion{Ca}{1} column densities of 3.4, 1.1, and 0.6 $\times$ 10$^9$~cm$^{-2}$, respectively.
For the seven lines of sight observed both with the coud\'{e} feed and at McDonald, the \ion{Ca}{1} equivalent widths agree within their mutual 2$\sigma$ uncertainties, with no apparent systematic differences. 
The absence of absorption features at the 0.2 m\AA\ level toward $\alpha$ Vir (observed at an airmass of about 1.5) suggests that there are no significant telluric lines near the \ion{Ca}{1} line.

\section{Results}
\label{sec-res}

\subsection{Spectra and Profile Analysis}
{\label{sec-spec}

The normalized \ion{Ca}{1} spectra are shown in Figure~1.
For lines of sight observed both with the coud\'{e} feed and at higher resolution, the coud\'{e} feed spectrum is shown at the top, with the McDonald or UHRF spectrum offset below. 
In most cases, the vertical scale has been expanded by a factor of 5 or 10 to show more of the structure in these generally very weak \ion{Ca}{1} lines.
Below \ion{Ca}{1} in each panel, we show the corresponding high resolution \ion{K}{1} (or \ion{Na}{1} D1) and \ion{Ca}{2} spectra, generally from WH, WHK, and WMH.
The source of each spectrum is indicated at the right, just above the continuum.
We have chosen the \ion{K}{1} and \ion{Na}{1} D1 reference wavelengths to be the $f$-value-weighted means of the wavelengths of the individual hyperfine subcomponents (7698.974 \AA\ and 5895.9242 \AA) to facilitate comparisons among the profiles.
(Note the resolved \ion{Na}{1} hyperfine structure, with separation $\sim$ 1 km~s$^{-1}$, for several components toward $\epsilon$ Ori; the \ion{K}{1} hyperfine structure, with separation $\sim$ 0.3 km~s$^{-1}$, is not seen.)

We have used the method of profile fitting to determine column densities ($N$), line widths ($b$ $\sim$ FWHM/1.665), and velocities ($v$) for the discernible individual interstellar clouds contributing to the observed line profiles.
We used Gaussian instrumental profiles, with the FWHM noted above, to fit the profiles observed with the different spectrographs.
The \ion{Ca}{1} rest wavelength (4226.728 \AA\ in air) and oscillator strength (1.75) were taken from the compilation of Morton (1991). 
In general, we adopted the minimum number of components needed to adequately fit the observed profiles, given the S/N achieved in each case --- such that the rms deviation between the data and the fitted profile is comparable to that previously measured for the adjacent continuum.
Components were added, as needed, to account for apparent asymmetries and/or local systematic deviations from the initial simpler fits.
In some cases, we used comparisons with similar high-resolution spectra of the stronger lines of \ion{K}{1}, \ion{Ca}{2}, and/or \ion{Fe}{1} to infer more complex structure (e.g., for the main absorption toward $\zeta$ Per and $\zeta$ Oph) or to confirm very weak components of marginal statistical significance in the \ion{Ca}{1} spectra (e.g., toward $\epsilon$ Ori, $\alpha$ Cyg, and a few other stars).

To illustrate the fitting procedure, in Figure~\ref{fig:fit} we show the individual components (dotted lines) and the composite fitted profiles (smooth solid lines) for two-component and four-component fits to the \ion{Ca}{1} absorption observed toward o Per (solid histograms).
The residuals (data minus fit) are shown as the lower histogram in each panel.
The two-component fit to this profile (with velocities near 11.3 and 13.6 km~s$^{-1}$) is inadequate, as the rms deviation between the data and fit within the absorption-line region (0.0071) is noticeably larger than the value observed in the nearby continuum (0.0062). 
Moreover, the observed absorption near both those velocities appears asymmetric; single-component fits to the left and right halves of the profile yield rms deviations of 0.0071 and 0.0065, respectively.
Splitting either of those two components into two sub-components reduces the overall fit rms deviation to 0.0056 or 0.0059 --- i.e., below the continuum rms deviation --- but does not address the apparent asymmetry in the other component.
Four components (at minimum) appear necessary, and yield a fit rms deviation of 0.0047.
Comparison with the \ion{K}{1} spectrum (Fig.~\ref{fig:spec1}; observed at comparably high resolution) indicates that four components --- at very similar velocities to those found for \ion{Ca}{1} --- are also needed to fit the strongest \ion{K}{1} absorption (WH).

The individual components derived in the fits to all the \ion{Ca}{1} profiles are noted in Figure~1 by tick marks above the spectra.
The derived individual component parameters are listed in Table~\ref{tab:comp}, where successive columns give the star name, observatory, component number, heliocentric velocity, column density, and $b$-value.
Of the 112 individual interstellar cloud components listed in Table~\ref{tab:comp}, 70 were derived from UHRF and/or McDonald spectra (15 stars), with the remaining 42 from the lower resolution coud\'{e} feed spectra (15 stars).
Where we have spectra from both the coud\'{e} feed and McDonald, the component parameters are those derived from the latter (higher resolution) data.

As was found in our previous high-resolution surveys, these new higher-resolution, higher S/N \ion{Ca}{1} spectra show complex component structure in most lines of sight. 
Where we have very high resolution spectra (FWHM $\la$ 0.6 km~s$^{-1}$) for \ion{Ca}{1} and for the stronger lines of \ion{K}{1}, \ion{Na}{1}, and/or \ion{Ca}{2}, the component velocity structures derived from independent fits to the various line profiles are generally very similar.
The \ion{Ca}{1} components generally correspond to the stronger components seen in \ion{K}{1} and/or \ion{Ca}{2}.
Where we have \ion{Ca}{1} spectra only from the coud\'{e} feed (lower resolution and often lower S/N), comparisons with the corresponding higher resolution spectra of the stronger lines suggest that the true \ion{Ca}{1} structure is more complex than we have been able to discern.

The profile fits also yield estimates for the uncertainties in the component parameters.
For well-defined components, the formal 1$\sigma$ uncertainties in the column densities are typically no more than 10--20\% (but at least several times $10^8$ cm$^{-2}$) --- comparable to the uncertainties that would be inferred from the error bars on the equivalent widths.
The uncertainties in the column densities can be somewhat larger for significantly blended components, however.
For most well-defined components in the McDonald and UHRF data, the derived velocities have formal uncertainties less than 0.1 km~s$^{-1}$, but the uncertainties are generally somewhat larger for broader or more severely blended components and for velocities derived from the coud\'{e} feed spectra.
Comparisons with the \ion{K}{1} spectra suggest that the absolute velocities for most well-defined components should be accurate to within about 0.2--0.3 km~s$^{-1}$.
The formal uncertainties in the line widths are typically of order 20\% for well-defined components in the higher-resolution spectra, but can be 30--50\% for weaker, narrower ($b$ $<$ 0.5 km~s$^{-1}$), and/or more blended components.
The nine $b$-values enclosed in square brackets were fixed to facilitate convergence of the fits, and are about as well determined as the $b$-values allowed to vary.
The 42 $b$-values enclosed in parentheses, often for weak and/or blended components, were essentially arbitrarily (though not unreasonably) fixed in the fitting, and are less well determined.
As for \ion{Na}{1}, \ion{Ca}{2}, and \ion{K}{1}, uncertainties in the component parameters due to not discerning the full \ion{Ca}{1} component structure may well be more significant than the formal uncertainties derived in the profile fits.

\subsection{Comparisons with Other Ca I Data}
\label{sec-comphi}
 
Because the \ion{Ca}{1} lines are typically very weak, very few high resolution (FWHM $\la$ 2 km~s$^{-1}$) spectra had previously been obtained (Hobbs 1971, using the PEPSIOS interferometric spectrometer); 
in many cases, only equivalent widths were reported (e.g., White 1973; Lambert \& Danks 1986; Gredel, van Dishoeck, \& Black 1993; Allen 1994).
(Note that for the last three references, the primary focus was the nearby CH$^+$ line at 4232 \AA, so any discussion of \ion{Ca}{1} was rather brief.)
In the last column of Table~\ref{tab:stel}, we list equivalent widths from various prior studies for the stars in our current sample.
In most cases, the values measured for a given line of sight are consistent within the listed 2$\sigma$ uncertainties.
Some of the upper limits reported by Hobbs (1971), computed for single, unresolved absorption features, are lower than subsequent detections, however. 
In all such cases, at least two weak components are detected in the new spectra.

Comparisons of the new high-resolution \ion{Ca}{1} spectra with those shown by Hobbs (1971) and Federman \& Hobbs (1983) reveal apparent differences in several cases.  
(1) Toward $\lambda$ Ori, Federman \& Hobbs found two components of equal equivalent width [1.9$\pm$0.4 m\AA\ (2$\sigma$)], at heliocentric velocities of about 6.6 and 24.2 km~s$^{-1}$.
While the higher velocity feature is reasonably consistent with the main absorption shown in Figure~\ref{fig:spec2}, the lower velocity feature is much stronger than the weak component near 7.6 km~s$^{-1}$ seen (tentatively) in the new McDonald spectrum.
There is no evidence for significant temporal variations in the stronger \ion{Na}{1} D lines at that velocity, over a somewhat longer time period (Hobbs 1969; Watson 1998; Price et al. 2001).
(2) Toward 23 Ori, roughly 30\% of the absorption shown by Federman \& Hobbs (of order 1.5 m\AA) appears to lie between about 12 and 16 km~s$^{-1}$, but none is seen at those velocities in the coud\'{e} feed spectrum in Figure~\ref{fig:spec2}.
While the absorption features toward $\lambda$ Ori and 23 Ori appear to be statistically significant in the spectra shown by Federman \& Hobbs, we note that there are comparable, but apparently spurious features (both positive and negative) toward several other stars in their small sample.
(3) The new McDonald spectrum of $\lambda$ Cep (Fig.~\ref{fig:spec5}) shows a fairly strong, narrow \ion{Ca}{1} component near $-$3 km~s$^{-1}$. 
While that component is not apparent in the spectrum shown by Hobbs (1971), that velocity is near the end of the PEPSIOS scan, and better agreement between the spectra might be obtained by adjusting the continuum in the earlier data.

\section{Discussion}
\label{sec-disc}

\subsection{Component Statistics}
\label{sec-stat}

In our previous surveys of \ion{Na}{1}, \ion{Ca}{2}, and \ion{K}{1} (WHK; WMH; WH), we examined the statistical properties of the ensembles of individual component column densities, line widths, velocities, and velocity separations between adjacent components ($\delta v$).
For all three species, the true median $\delta v$ (allowing for additional unresolved structure) is estimated to be of order 1.2 km~s$^{-1}$.
The median line widths for individual \ion{Na}{1} and \ion{K}{1} components are both about 1.1--1.2 km~s$^{-1}$ (FWHM); the corresponding \ion{Ca}{2} components are typically somewhat broader, however.
Adjacent components can have rather different properties [e.g., line widths, $N$(\ion{Na}{1})/$N$(\ion{Ca}{2}) ratios].
Those references also discuss the effects of limited samples, inhomogeneities in resolution and S/N, and the likelihood that we have not discerned all the components actually present.

Because the \ion{Ca}{1} absorption is generally weak, the number of components detected in this survey is significantly smaller, and provides little new information on the statistics of $v$, $\delta v$, and $N$;
we merely note that the median individual component \ion{Ca}{1} column density is of order 2 $\times$ 10$^9$ cm$^{-2}$.
While the set of \ion{Ca}{1} component $b$-values is rather limited, however, it does provide interesting constraints on the temperature, internal turbulent velocity ($v_t$), and distribution of the gas traced by \ion{Ca}{1}, since
$b$ = $(2kT/m + 2v_t^2)^{1/2}$, where $m$ is the atomic weight.\footnotemark\
The median $b$ for the primary sample of 47 well determined $b$-values derived from the high-resolution UHRF and/or McDonald spectra is 0.66 km~s$^{-1}$, which corresponds to a maximum temperature $T_{\rm max}$ $\sim$ 1030 K or to a maximum turbulent velocity $v_{t~{\rm max}}$ $\sim$ 0.47 km~s$^{-1}$.
Roughly 85\% of the primary $b$-values fall between 0.2 and 0.9 km~s$^{-1}$, 
corresponding to values of $T_{\rm max}$ between 95 and 1920 K or values of 
$v_{t~{\rm max}}$ between 0.14 and 0.64 km~s$^{-1}$.
The median $b$-values for the coud\'{e} feed and combined samples are both somewhat larger, reflecting the less complete knowledge of the component structures in the lower resolution spectra.
Besides thermal and turbulent broadening, unresolved blends and cloud stratification may also contribute to the widths of the observed \ion{Ca}{1} lines (see WHK and WMH).
Comparisons with the profiles of \ion{K}{1} and \ion{Ca}{2} do suggest that most \ion{Ca}{1} components with $b$ $>$ 1 km~s$^{-1}$ may in fact consist of unresolved blends.
The ``true'' median \ion{Ca}{1} $b$-value therefore could be somewhat smaller.
\footnotetext{In this definition of $b$, $v_t$ is the one-dimensional rms value for a Gaussian spectrum of internal turbulent velocities; the full three-dimensional turbulent velocity is thus 3$^{1/2}v_t$.}

If \ion{Ca}{1}, \ion{Ca}{2} ($m$ = 40), and \ion{K}{1} ($m$ = 39) are similarly distributed, we would expect that the respective $b$-values would be essentially identical, regardless of whether thermal broadening or turbulent broadening is dominant.
For the components in the respective high resolution primary samples, the median $b$-values are $b_{\rm med}$(\ion{Ca}{1}) = 0.66 km~s$^{-1}$, $b_{\rm med}$(\ion{Ca}{2}) = 1.33 km~s$^{-1}$, and $b_{\rm med}$(\ion{K}{1}) = 0.67 km~s$^{-1}$.
We have identified 35 ``corresponding'' components in \ion{Ca}{1} and \ion{K}{1} for which the velocities agree to within 0.4 km~s$^{-1}$ and for which any immediately neighboring components have roughly similar properties (as for the corresponding \ion{Na}{1} and \ion{Ca}{2} components discussed by WMH).
The median values of $b$ are 0.69 and 0.65 km~s$^{-1}$ for the corresponding \ion{Ca}{1} and \ion{K}{1} components, respectively.
The close agreement for the \ion{Ca}{1} and \ion{K}{1} $b$-values suggests that those two species are essentially coextensive in the interstellar gas --- and also with \ion{Na}{1} (WH).
While the typical uncertainties in the $b$-values would allow $b$(\ion{Ca}{1}) $\sim$ $b$(\ion{Ca}{2}) for many of the corresponding \ion{Ca}{2} components with $b$ $\la$ 1.2 km~s$^{-1}$, the data suggest that $b$(\ion{Ca}{2}) is often greater than $b$(\ion{Ca}{1}) --- both globally and for individual components seen at the same velocity.
Barlow et al. (1995) and WMH have suggested that this difference in $b$-values may imply that \ion{Ca}{2} and the three trace neutral species are not identically distributed, even though similarities in component velocities suggest that the various species are associated.
For a given velocity component, \ion{Ca}{2} may occupy a somewhat larger volume, characterized by a larger temperature and/or greater internal turbulent motions (where Ca might be less severely depleted), than do the trace neutrals.

\subsection{Comparisons of Ca I with Other Species}
\label{sec-compar}
 
In this section, we examine some of the relationships between \ion{Ca}{1}, \ion{Ca}{2}, \ion{K}{1}, \ion{Na}{1}, \ion{C}{1}, \ion{S}{1}, \ion{Fe}{1}, H~(\ion{H}{1} + 2H$_2$), and H$_2$ --- which (in principle) can yield insights into the abundances, depletions, and physical conditions in the predominantly neutral clouds in which those species reside.
The interpretation of the observed relationships, however, involves several assumptions and approximations which are not strictly valid, and which can bias conclusions drawn from those relationships.
We begin by briefly examining those assumptions and noting how they may affect our interpretation of the observed data.

\subsubsection{Assumptions and Approximations}
\label{sec-approx}

While we would like to determine abundances and physical conditions at specific locations within individual interstellar clouds, we do not have the data to do so.
The high-resolution spectra available for some lines of sight suggest that most sightlines are complex, containing multiple blended components (clouds) which can be characterized by different relative abundances and physical properties.
For example, it is clear from the profiles in Figure~1 that \ion{Ca}{2} is often detected (and sometimes strong) over a much wider velocity range than \ion{Ca}{1} and \ion{K}{1}.
For many species and many sightlines, however, we have only total column densities integrated over the whole line of sight --- whether because of low spectral resolution (for most UV lines) or, for \ion{H}{1} and H$_2$, the inextricable blending of very strong, broad individual components.
Where we do have very high-resolution spectra, we may obtain more detailed information by comparing column densities for corresponding individual components identified in fits to the line profiles (e.g., Welty et al. 1999b) or, alternatively, by comparing ``apparent column densities'' as functions of velocity (e.g., Jenkins \& Tripp 2001).
Even then, however, it is likely that individual clouds are not homogeneous, and that different species may be distributed somewhat differently within the clouds.
For example, considerations of ionization equilibrium suggest that trace neutral species will be more concentrated in the denser parts of the clouds than the corresponding dominant first ions; differences in photoionization cross-section and depletion behavior may lead to further differences in distribution among the various trace and dominant species.
The generally larger $b$-values found for \ion{Ca}{2}, compared to those found for \ion{Ca}{1} and \ion{K}{1}, may reflect a combination of those effects.

It is commonly assumed that photoionization and radiative recombination dominate the ionization equilibrium of heavy elements in diffuse clouds, such that 
\begin{equation}
\Gamma({\rm X}^0)~n({\rm X}^0)~=~\alpha_r({\rm X}^0,T)~n_e~n({\rm X}^+), 
\end{equation}
where $\Gamma$ is the photoionization rate and $\alpha_r$ is the radiative recombination rate coefficient.
It is also usual to assume (despite the potential problems noted above) that the local volume densities ($n$) may be replaced by the corresponding column densities ($N$) --- either for individual components or (in most cases) for the total line of sight.
Ratios of trace neutral species with the corresponding dominant first ions [measured directly or estimated from $N$(H) and assumed depletions] thus can yield estimates for $n_e$ (if the temperature and radiation field can be constrained) and the local hydrogen density $n_{\rm H}$ (if the fractional ionization $n_e$/$n_{\rm H}$ is assumed to be equal to the gas phase abundance of carbon).

In the discussions below, we will use the equations of photoionization equilibrium, with the photoionization rates calculated for the WJ1 radiation field (de Boer et al. 1973) and the radiative recombination rates at $T$ = 100 K (usually), as tabulated by P\'{e}quignot \& Aldrovandi (1986).
We define the electron density derived under the assumption that radiative recombination dominates as 
$n_e$(rad) = ($\Gamma$/$\alpha_r$)[$N$(\ion{X}{1})/$N$(\ion{X}{2})].
Unless otherwise noted, we use total line-of-sight column densities, since individual component values generally are not available for hydrogen, \ion{C}{1}, or \ion{S}{1}.
Column densities for \ion{Ca}{1}, \ion{Ca}{2}, \ion{Fe}{1}, and \ion{S}{1} (with references and a brief discussion of uncertainties) are given in Appendix B of this paper (Table~\ref{tab:coldens}); values for the other species are given in the appendix to WH.
We also adopt the solar system meteoritic reference abundances (A$_{\sun}$) listed by Anders \& Grevesse (1989), except for carbon, for which we take the value for the solar photosphere from Grevesse \& Noels (1993)\footnotemark.
The reference abundances and rates are collected in Table~\ref{tab:atom}.
\footnotetext{Recent slight revisions for some of the reference abundances (e.g., Grevesse \& Sauval 1998; Holweger 2001) would have a minor impact on the results discussed in this paper.}

While we will adopt those assumptions (at least initially) in the discussions that follow, we will find in some cases that they lead to results that are inconsistent with those implied by other diagnostics of the physical conditions within the clouds --- suggesting that some (or all) of the assumptions may not be valid.
For example, the electron densities inferred from different $N$(\ion{X}{1})/$N$(\ion{X}{2}) ratios in a given sightline can differ by an order of magnitude or more (e.g., Fitzpatrick \& Spitzer 1997; Welty et al. 1999b).
Moreover, the estimates for $n_e$ and $n_{\rm H}$ can be systematically higher than the values derived from analyses of the fine-structure excitation of \ion{C}{2} and \ion{C}{1} (Welty et al. 1999b; WH).
Such systematic differences suggest that other processes besides photoionization and radiative recombination may have a significant effect on the ionization balance of heavy elements in at least some diffuse clouds.

One promising candidate for enhancing the various trace neutral species is charge exchange between the corresponding singly ionized species and neutral and/or negatively charged large molecules/small grains (Lepp et al. 1988; WH; Weingartner \& Draine 2001; Liszt 2003).
If this ``grain-assisted'' recombination dominates over radiative recombination, we have 
\begin{equation}
\Gamma({\rm X}^0)~n({\rm X}^0)~=~\alpha_g({\rm X}^0,\psi,T)~n_{\rm H}~n({\rm X}^+), 
\end{equation}
where $\alpha_g$ is the recombination rate coefficient ({\it per hydrogen nucleus}).
In this definition, $\alpha_g$ includes the effects of the grain abundance, the grain size distribution, and the grain charge distribution; $\alpha_g$ depends on $n_e$, $T$, and the radiation field through the ``charging parameter'' $\psi$ = $G \sqrt{T}/n_e$, where the energy density in the ambient interstellar radiation field between 6 and 13.6 eV is $G$ times that of the Habing (1968) field.
In the next-to-last column of Table~\ref{tab:atom}, we list values of $\alpha_g$ for the ``typical cold, neutral medium'' ($T$ = 100 K, $\psi$ = 400 K$^{1/2}$ cm$^3$), calculated via equation 8 in Weingartner \& Draine (2001).

Equation 16 of Weingartner \& Draine (2001) gives the electron density for the more general case in which both radiative and grain-assisted recombination may be significant:
\begin{equation}
n_e = \frac{R~\Gamma}{\alpha_r}
\left[1 - \left(\frac{1-s+R}{1+R}\right)~\frac{\alpha_g~n_{\rm H}}{R~\Gamma}\right],
\end{equation}
where $R$ = $N$(\ion{X}{1})/$N$(\ion{X}{2}) is usually much less than 1.0, $s$ is the probability that an ion sticks to the grain after charge exchange, and the quantity in front of the square braces is $n_e$(rad). 
For any assumed $s$, the effect of including grain-assisted recombination thus depends in a somewhat complicated way on the column density ratio $R$, the local density $n_{\rm H}$, the temperature (mildly), and the strength of the radiation field --- directly and/or through the dependence of $\alpha_g$ on those factors.
In general, the effect is largest for small $R$ (and thus usually for small inferred $n_e$), for large $n_{\rm H}$, and for small $\Gamma$/$\alpha_g$ (e.g., for Ca$^+$, Na, Mg, and K).
For $s$ = 1 (i.e., all the neutralized ions are incorporated into the grains, and not returned to the gas phase), the quantity in the square braces becomes 
[1 $-$ $\alpha_g$ $n_{\rm H}$ / $\Gamma$] $\sim$ 1 (unless $n_{\rm H}$ is much greater than 10$^3$ cm$^{-3}$), so that $n_e$ $\sim$ $n_e$(rad).
For $s$ = 0 (which maximizes the effect of the grain-assisted recombination), the above equation can be re-arranged to yield
$R$ = $n_e$(rad) [1 + ($\alpha_g$ $n_{\rm H}$)/($\alpha_r$ $n_e$)].
The ratio ($\alpha_g$ $n_{\rm H}$)/($\alpha_r$ $n_e$), effectively a measure of the relative contributions of grain-assisted and radiative recombination, can range from about 0.1--10 for fractional ionizations between 2 $\times$ 10$^{-4}$ and 0.01, for the recombination coefficients listed in Table~\ref{tab:atom}.
The effect of grain-assisted recombination thus can vary for different elements in any given line of sight (depending on $\alpha_g$/$\alpha_r$ and $s$), and can also vary from one sightline to another for any given element (depending on $R$, $T$, $\Gamma$, and $n_{\rm H}$).

In the following sections, we examine a number of possible relationships between individual ions and also between ratios of different atomic and molecular species, in an attempt to determine which physical processes have an important effect on the abundances of the various species.
In each case, we first determine whether or not a correlation exists between the two quantities, using several different estimates for the correlation probability.
The first estimate is that associated with the linear correlation coefficient, which is derived from the values of all points detected in both variables (e.g., Bevington 1969).
As we have chosen to weight the data (using 1/$\sigma^2$), the tabulated linear correlation coefficients and corresponding probabilities are averages of the values determined for considering each variable as the dependent one.
Because only limits (for one or both of the variables) are available for many of the points in some of the possible correlations, we also provide estimates for the correlation probabilities using the three bivariate correlation tests in the survival analysis (or censored data analysis) package ASURV (Isobe, Feigelson, \& Nelson 1986), which use both detections and limits.\footnotemark\
\footnotetext{Specifically, we used Cox's proportional hazard model, the generalized Kendall's tau method, and the generalized Spearman's rank order method.
Version 1.2 of the ASURV package (LaValley, Isobe, \& Feigelson 1992) was obtained from the statistical software archive maintained at Penn State (http://www.astro.psu.edu/statcodes).
These routines do not allow weighting of the data.}
(Note that in all cases we give the probabilities of {\it no} correlation --- i.e., a small value for the probability corresponds to a strong correlation.)
While the four different methods yield concordant results for most of the possible correlations, there are several cases for which the survival analysis tests suggest that a correlation is present, even though the linear correlation coefficient suggested otherwise.
For the confirmed correlations, we then perform a regression analysis which allows for uncertainties in both variables (but which uses only the detected points), and which yields both the slope and intercept of the ``typical'' relationship and an estimate of the scatter (WH).\footnotemark\
\footnotetext{We used a slightly modified version of the subroutine regrwt.f, obtained from the Penn State statistical software archive, to perform the regression fits.
While WH assumed equal weights, here we weight the data by 1/$\sigma^2$; the weighted and unweighted fits generally yield similar results, however.
The bivariate regression routines in survival analysis packages like ASURV do not use uncertainties (for either variable), and cannot be used when there are both upper and lower limits for either variable.}
For some of the relationships, we also perform the fits with the slope fixed at 1.0 or 0.0 for easier comparison with theoretical approximations.

As much as possible, we attempt to minimize biases resulting from differences in the spatial distribution of different species (as noted above) by comparing or considering ratios of species that are likely to be similarly distributed, by virtue of having similar dependences on temperature, density, and/or ambient radiation field.
For example, both considerations of ionization equilibrium and empirical evidence suggest that the various trace neutral species (e.g., \ion{C}{1}, \ion{Na}{1}, \ion{K}{1}) are fairly well correlated with each other --- though there may be some differences in ionization and/or depletion (as discussed below).
Moreover, the column densities of both \ion{Na}{1} and \ion{K}{1} exhibit fairly well-defined relationships with the column densities of the molecular species CH and H$_2$ (with some exceptions; WH).
Most dominant species (e.g., \ion{Si}{2}, \ion{Fe}{2}, \ion{Zn}{2}) should also be fairly well correlated with each other, though (again) there will be differences due to depletion behavior (see, for example, the absorption-line profiles and curves of growth for the line of sight to 23 Ori in Welty et al. 1999b).
Even \ion{Ca}{2}, whose behavior exhibits an interplay between ionization and depletion effects, shows a surprisingly good correlation with \ion{Fe}{2}, which is typically a ``dominant-depleted'' species in the primarily neutral clouds sampled in this study (Welty et al. 1996, 1999a, 1999b).
 
For example, where column densities for several trace neutral species are available, we may estimate {\it relative} abundances/depletions for those elements by taking ratios of the ionization equilibrium equation:
\begin{equation}
\frac{\delta({\rm X}){\rm A}_{\odot}({\rm X})}{\delta({\rm Y}){\rm A}_{\odot}({\rm Y})} \sim
\frac{N({\rm X~II})}{N({\rm Y~II})} =
\frac{(\Gamma/\alpha)_{\rm X}}{(\Gamma/\alpha)_{\rm Y}}~
\frac{N({\rm X~I})}{N({\rm Y~I})},
\end{equation}
where $\delta$(X) is the depletion of X, A$_{\odot}$ is the solar reference abundance of X, and we have assumed that the singly ionized species are dominant.
In taking such ratios, the dependences on the density ($n_e$ and/or $n_{\rm H}$), on the temperature [through $\alpha(T)$] and on the radiation field (through $\Gamma$) --- all of which can be rather uncertain --- are substantially reduced (York 1980; Snow 1984).
If radiative recombination dominates the production of the trace neutral species, then the effective recombination rate coefficient $\alpha$ would be equal to $\alpha_r$; if charge exchange with grains dominates, then $\alpha$ $\sim$ $\alpha_g$.
If both processes are significant, then $\alpha$ would be replaced by $\alpha_r$[1 + ($\alpha_g$ $n_{\rm H}$)/($\alpha_r$ $n_e$)] or (equivalently) by
$\alpha_g$[1 + ($\alpha_r$ $n_e$)/($\alpha_g$ $n_{\rm H}$)] (for $s$ = 0; see above).
The ratios $n$(\ion{X}{2})/$n$(\ion{Y}{2}) and $n$(\ion{X}{1})/$n$(\ion{Y}{1}) should vary much less through each cloud than either the individual densities or the trace/dominant ratios.
Because the depletion of carbon in the local ISM is apparently uniform [$\delta$(C) $\sim$ 0.4, relative to solar abundances (Cardelli et al. 1996; Sofia et al. 1997)], the roughly constant ratios of the column densities of \ion{Li}{1}, \ion{Na}{1}, and \ion{K}{1}, relative to that of \ion{C}{1}, have thus yielded values for the typical depletions of lithium, sodium, and potassium:  D(X) = log $\delta$(X) $\sim$ $-$0.6, $-$0.6, and $-$0.7 dex, respectively, for $\alpha$ $\sim$ $\alpha_r$ (WH).

\subsubsection{Calcium Ionization Balance}
\label{sec-ionbal}

Measurements for two adjacent ionization states (\ion{Ca}{1} and \ion{Ca}{2}) yield constraints on the ionization of calcium in interstellar clouds.
Because the ionization potential of \ion{Ca}{2} is about 11.87 eV, \ion{Ca}{3} is expected to be the dominant form of calcium in very diffuse clouds. 
The combination of a relatively small photoionization rate for \ion{Ca}{2} and a relatively large radiative recombination rate coefficient for \ion{Ca}{3} (Table~\ref{tab:atom}), however, suggests that \ion{Ca}{2} could be dominant in somewhat thicker clouds.
Taking the equilibrium equations for two successive ionization states, 
$N$(\ion{Ca}{1})/$N$(\ion{Ca}{2}) = $n_e$/($\Gamma$/$\alpha_r$)$_{\rm Ca I}$ and 
$N$(\ion{Ca}{2})/$N$(\ion{Ca}{3}) = $n_e$/($\Gamma$/$\alpha_r$)$_{\rm Ca II}$, 
we obtain 
$N$(\ion{Ca}{2})/$N$(\ion{Ca}{3}) $\sim$ 1300 $N$(\ion{Ca}{1})/$N$(\ion{Ca}{2}).
For individual components with $N$(\ion{Ca}{1})/$N$(\ion{Ca}{2}) in the observed range 0.0015-0.08, we thus have $N$(\ion{Ca}{2})/$N$(\ion{Ca}{3}) $\sim$ 2--100 --- so that \ion{Ca}{2} would be the dominant form of calcium wherever \ion{Ca}{1} is detected.

Several additional factors may modify that result, however.
(1) If \ion{Ca}{2} is more broadly distributed than \ion{Ca}{1} within individual clouds (as discussed above), then the $N$(\ion{Ca}{2})/$N$(\ion{Ca}{3}) ratio would be even larger in the region occupied by \ion{Ca}{1}.
The magnitude of this effect is difficult to estimate, however.
(2) On the other hand, if the recombination is dominated by charge exchange with small grains [or large molecules, with $\alpha_g$(\ion{Ca}{3}) $\sim$ 3 $\times$ $\alpha_g$(\ion{Ca}{2}) (Weingartner \& Draine 2001)], then $N$(\ion{Ca}{2})/$N$(\ion{Ca}{3}) would be reduced by a factor of about 1.7, and \ion{Ca}{3} would be comparable to \ion{Ca}{2} in cases where $N$(\ion{Ca}{1})/$N$(\ion{Ca}{2}) is low.
(3) Moreover, if the unattenuated ($\Gamma$/$\alpha_r$)$_{\rm Ca I}$ is actually smaller than the value listed in Table~\ref{tab:atom} (a possibility noted below), then $N$(\ion{Ca}{2})/$N$(\ion{Ca}{3}) would be correspondingly reduced, implying more (unobserved) \ion{Ca}{3}, and thus less severe depletion of calcium than inferred from \ion{Ca}{2} alone [again, where $N$(\ion{Ca}{1})/$N$(\ion{Ca}{2}) is low].

\subsubsection{Ca I (and Other Trace Neutral Species) vs. K I}
\label{sec-ca1k1}

Welty \& Hobbs (2001) found fairly tight, essentially linear relationships between the logarithmic column densities of \ion{Li}{1}, \ion{Na}{1}, and \ion{K}{1}, with slopes near unity and rms scatter less than about 0.2 dex.
The close correspondence between those three species suggests that they respond very similarly to changes in physical conditions in the ISM and that they are generally coextensive --- which is not surprising, in view of similarities in ionization cross-section and chemical behavior.
(Slight differences were seen, particularly for \ion{Na}{1}, for sightlines in the Sco-Oph region and in several other regions, however.)
While the column density of \ion{C}{1} also showed a strong correlation with the column densities of \ion{Na}{1} and \ion{K}{1}, the slightly steeper slopes (1.1--1.3) may reflect the differential effects of extinction on the various photoionization rates in the higher column density sightlines (see below) and/or uncertainties in the \ion{C}{1} $f$-values (Jenkins \& Shaya 1979; WH; Jenkins \& Tripp 2001).

The additional data collected in Appendix B of this paper permit similar comparisons between $N$(\ion{K}{1}) and the column densities of \ion{Ca}{1}, \ion{S}{1}, and \ion{Fe}{1}.
For example, in Figure~\ref{fig:ca1k1} we plot the total sightline values of log[$N$(\ion{Ca}{1})] vs. log[$N$(\ion{K}{1})].
While the correlation probabilities listed in Table~\ref{tab:corr} indicate strong correlations in all three cases, the fitted regression slopes (Table~\ref{tab:fits}) all differ from unity.
The slope of the best-fit (solid) line for log[$N$(\ion{Ca}{1})] vs. log[$N$(\ion{K}{1})] is 0.60$\pm$0.06 (Table~\ref{tab:fits}; all sightlines with detections of both species). 
Depending on the sample used (including the Sco-Oph sightlines or not), the slopes for log[$N$(\ion{S}{1})] and log[$N$(\ion{Fe}{1})] are about 1.2--1.5 and 0.8--0.9, respectively.
The rms scatter for the unconstrained fits, in all cases between 0.12 and 0.21 dex, is somewhat larger than the values expected from the nominal uncertainties on the individual column densities.
Either the uncertainties have been underestimated or there are real deviations from the mean relationships in some sightlines.
The slight reductions in the scatter when the Sco-Oph sightlines are excluded and the somewhat larger scatter for the correlations involving \ion{Ca}{1} and/or \ion{Fe}{1} (which are expected to be more affected by variations in depletion) both suggest that real deviations are present.

In principle, several effects could lead to slopes greater than or less than unity in comparisons between log[$N$(\ion{X}{1})] and log[$N$(\ion{K}{1})].
One possibility is that of differential extinction effects (noted above for \ion{C}{1}).
Differences in photoionization cross-section for the various neutral species, coupled with the wavelength dependences of both the interstellar radiation field and the extinction due to dust, imply that ratios of the photoionization rates [$\Gamma$(X$^0$)/$\Gamma$(Y$^0$)] will vary somewhat with depth within individual interstellar clouds (e.g., Roberge, Dalgarno, \& Flannery 1981).
In column 5 of Table~\ref{tab:atom}, we give the change in log($\Gamma$) between the edge (the unattenuated value) and the center of a cloud with total visual extinction 1 mag, for grain model 2 of Roberge et al. (drawn from their Figure~2b).
Those values suggest that \ion{C}{1}, \ion{S}{1}, and \ion{Ca}{2} (with the highest ionization potentials) would be slightly enhanced (by 0.15--0.24 dex), relative to \ion{K}{1}, in going from the edge to the center of the cloud, while species such as \ion{Mg}{1}, \ion{Al}{1}, and \ion{Si}{1} would be slightly reduced (by 0.10--0.14 dex).
The relative reductions of \ion{Na}{1}, \ion{Ca}{1}, and \ion{Fe}{1} would be even smaller ($<$ 0.1 dex).
The relationships between $N$(\ion{K}{1}), $N$(H), and $E(B-V)$ discussed by WH suggest, however, that even the thickest (highest column density) individual clouds in our sample probably have total visual extinctions of only about 0.6 mag --- so that the effects of differential extinction would be even smaller than those estimated above (especially when integrated over the whole cloud).
While the slightly steeper slopes found for \ion{C}{1} and \ion{S}{1} thus might be due (at least in part) to differential extinction in the thicker clouds, it seems unlikely that the shallower slopes found for \ion{Ca}{1} and \ion{Fe}{1} could be explained in that way.

A second possibility is that of differences in how the recombination rates of the various singly ionized species might vary --- either with depth within an individual cloud or between more diffuse and denser clouds.
As an example, we consider the ratio $N$(\ion{S}{1})/$N$(\ion{K}{1}), including radiative and/or grain-assisted recombination, for $T$ = 100 K and $\psi$ = 400 K$^{1/2}$ cm$^3$ (the ``typical'' values) and for $T$ = 25 K and $\psi$ = 100 K$^{1/2}$ cm$^3$ (representing colder, denser regions).
We choose those two species for several reasons:  
(1) differences in depletion are unlikely to be significant, 
(2) there are (slight) differences in the temperature dependence of $\alpha_r$ (P\'{e}quignot \& Aldrovandi 1986), 
(3) there are differences in the ratio $\alpha_r$/$\alpha_g$, and
(4) the observed relationship is steeper than linear.
If radiative recombination dominates, the ratio $\alpha_r$(S)/$\alpha_r$(K) is equal to 1.91 for the ``typical'' case and to 1.69 for the ``colder, denser'' case --- a difference of 13\% (0.05 dex).
If grain-assisted recombination dominates, the ratio $\alpha_g$(S)/$\alpha_g$(K) equals 1.36 and 1.44 for the two cases --- a difference of 6\% (0.02 dex).
When both radiative and grain-assisted recombination are significant, we take ratios of the quantity 
$\alpha_g$[1 + ($\alpha_r$ $n_e$)/($\alpha_g$ $n_{\rm H}$)] (for $s$ = 0; see \S~\ref{sec-approx} and \S~\ref{sec-relne}) for \ion{S}{1} with respect to its value for \ion{K}{1}; we further assume a constant fractional ionization $n_e$/$n_{\rm H}$ = 5 $\times$ 10$^{-4}$.
For the ``typical'' case, the ratio is equal to 1.50; for the ``colder, denser'' case, the ratio is equal to 1.53 --- only 2\% larger.
If the fractional ionization decreases at higher densities (a possibility noted below), the ``colder, denser'' ratio will decrease slightly, to a minimum of 1.44 (the ratio of the $\alpha_g$).
These rather slight differences suggest that differences in recombination rate behavior are not responsible for the observed steeper than linear relationship between $N$(\ion{S}{1}) and $N$(\ion{K}{1}).

A third possibility is that of differences in depletion behavior (e.g., with depth or density) for the different elements.
Observed increases in the severity of depletions (e.g., for iron, silicon, magnesium, calcium) for sightlines with higher mean densities ($<n_{\rm H}>$) or higher fractions of hydrogen in molecular form $f$(H$_2$) have suggested that the depletions may depend on the local hydrogen density $n_{\rm H}$ (e.g., Jenkins 1987; Cardelli 1994).
Wakker \& Mathis (2000) have shown similar trends for the depletions of several elements versus $N$(H).
Given the roughly quadratic dependence of $N$(\ion{K}{1}) on $N$(H) and the apparent lack of significant variation in the potassium depletion (WH), we thus would expect (1) that the ratio $N$(\ion{X}{1})/$N$(\ion{K}{1}) --- which is roughly proportional to the depletion of X (see next section) --- would decrease at higher overall column densities, and (2) that the slope of the relationship between log[$N$(\ion{X}{1})] and log[$N$(\ion{K}{1})] would be less than 1.0 for elements X which can be severely depleted.
For example, Wakker \& Mathis find that $N$(\ion{Fe}{2})/$N$(H) is roughly proportional to [$N$(H)]$^{-0.6}$, for 10$^{18.5}$ cm$^{-2}$ $\la$ $N$(H) $\la$ 10$^{21.5}$ cm$^{-2}$.
If $N$(\ion{K}{1}) is proportional to [$N$(H)]$^2$, then $N$(\ion{Fe}{1})/$N$(\ion{K}{1}) $\propto$ $N$(\ion{Fe}{2})/$N$(\ion{K}{2}) would be roughly proportional to [$N$(\ion{K}{1})]$^{-0.3}$, so that $N$(\ion{Fe}{1}) would be proportional to [$N$(\ion{K}{1})]$^{0.7}$ --- slightly weaker than the observed dependence.
Differences in depletion behavior thus can account quite naturally for the shallower slopes found for $N$(\ion{Ca}{1}) and $N$(\ion{Fe}{1}) [and also for \ion{Mg}{1}, for which fewer data are available], versus $N$(\ion{K}{1}).

\subsubsection{Ca I/K I and Ca I/Ca II vs. $f$(H$_2$)}
\label{sec-dep}

As noted above, the components detected in \ion{Ca}{1} generally have counterparts in both \ion{K}{1} and \ion{Ca}{2}. 
For \ion{Ca}{1} and \ion{K}{1}, the generally good correspondence in both velocity extent (Figure~1) and component $b$-values (\S~\ref{sec-stat}) suggests that the two species are essentially coextensive.
The typically larger $b$-values for \ion{Ca}{2}, however, suggest that \ion{Ca}{2} is somewhat more broadly distributed; moreover, \ion{Ca}{2} is often detected over a wider velocity range.
For the set of individual components detected in all three species, the column density ratios $N$(\ion{Ca}{1})/$N$(\ion{K}{1}) and $N$(\ion{Ca}{1})/$N$(\ion{Ca}{2}) can range over factors of about 500 and 50, respectively --- though the ranges for components within any individual sightline are generally much smaller than those extreme values.
The total line-of-sight values (or limits) for $N$(\ion{Ca}{1})/$N$(\ion{Ca}{2}) will underestimate the ``true'' values (i.e., for the components detected in \ion{Ca}{1}) for cases where \ion{Ca}{2} is detected over a much wider velocity range --- but generally not by more than a factor of 2.
In this section, we use the equations of photoionization equilibrium to interpret the behavior of those two column density ratios in the different environments probed by the sightlines in our sample. 

In photoionization equilibrium, the ratio 
\begin{equation}
\frac{N({\rm Ca~I})}{N({\rm K~I})} \sim
\frac{(\Gamma/\alpha_r)_{\rm K}}{(\Gamma/\alpha_r)_{\rm Ca}} 
\frac{[\delta({\rm Ca}) {\rm A}_{\odot}({\rm Ca})]}{[\delta({\rm K}) {\rm A}_{\odot}({\rm K})} \sim 12.4~\delta({\rm Ca}) 
\end{equation}
is sensitive primarily to the calcium depletion, with only weak dependence on $T$ or the radiation field.
We have assumed a constant potassium depletion [$\delta$(K) = 0.2] and that \ion{Ca}{2} and \ion{K}{2} are the dominant ionization states.
The ratio $N$(\ion{Ca}{1})/$N$(\ion{Ca}{2}) = $n_e$/($\Gamma$/$\alpha_r$) depends on the electron density, the ambient radiation field, and the temperature (since $\alpha_r$ $\sim$ $\alpha_0$ $T^{-\beta}$, with $\beta$ typically 0.6--0.7).
If the fractional ionization $n_e$/$n_{\rm H}$ is assumed constant, then the latter ratio would also depend on $n_{\rm H}$.
If charge exchange with grains were to dominate over radiative recombination, then $N$(\ion{Ca}{1})/$N$(\ion{K}{1}) $\sim$ 7.6~$\delta$(Ca) [with $\delta$(K) = 0.36] and 
$N$(\ion{Ca}{1})/$N$(\ion{Ca}{2}) = $n_{\rm H}$/($\Gamma$/$\alpha_g$).
If the sticking parameter $s$ $\sim$ 1 for calcium (typically severely depleted), however, then radiative recombination would be responsible for most of the \ion{Ca}{1}, and we could have 
\begin{equation}
\frac{N({\rm Ca~I})}{N({\rm K~I})} \sim
\frac{(\Gamma/\alpha_g)_{\rm K}}{(\Gamma/\alpha_r)_{\rm Ca}}~
\frac{n_e}{n_{\rm H}}~
\frac{[\delta({\rm Ca}) {\rm A}_{\odot}({\rm Ca})]}{[\delta({\rm K}) {\rm A}_{\odot}({\rm K})} \sim 
5 \times 10^3~\frac{n_e}{n_{\rm H}}~\delta({\rm Ca}).
\end{equation}

In Figure~\ref{fig:cakvsfh2}, we compare the two ratios $N$(\ion{Ca}{1})/$N$(\ion{K}{1}) and $N$(\ion{Ca}{1})/$N$(\ion{Ca}{2}) with the fraction of hydrogen in molecular form $f$(H$_2$), which is often considered to be indicative of the local density $n_{\rm H}$ (e.g., Cardelli 1994).
In this figure, the filled circles represent the total line-of-sight values, while the crosses represent individual components (only those detected in the relevant species). 
Sightlines in the Sco-Oph and Orion regions, some of which can exhibit deviations from the otherwise well-defined mean relationships in some comparisons (WH), are marked with open circles and open squares, respectively.
The individual components are assumed to have the same $f$(H$_2$) as the sightline as a whole --- which is likely a reasonable approximation for sightlines with low overall $f$(H$_2$), but not for minor components in sightlines with high $f$(H$_2$).
For higher densities, we might expect lower temperatures, higher molecular fractions, and higher $n_e$/($\Gamma$/$\alpha_r$) ratios.
For stronger radiation fields, we expect lower $f$(H$_2$).

The top plot, log[$N$(\ion{Ca}{1})/$N$(\ion{K}{1})] versus log[$f$(H$_2$)], indicates that $N$(\ion{Ca}{1})/$N$(\ion{K}{1}) is roughly constant for $f$(H$_2$) $\la$ 0.1, but decreases markedly for $f$(H$_2$) $\ga$ 0.1.
This behavior is very similar to that found for the gas phase abundances of dominant ions of elements whose depletions become more severe in colder, denser clouds [e.g., $N$(\ion{Fe}{2})/$N$(H)] --- consistent with the expectation that $N$(\ion{Ca}{1})/$N$(\ion{K}{1}) is an indicator of the depletion of calcium.

In the bottom plot, however, there is no apparent dependence of $N$(\ion{Ca}{1})/$N$(\ion{Ca}{2}) on $f$(H$_2$) --- suggesting either that $n_e$/($\Gamma$/$\alpha_r$) does not depend on the local density (see also \S~\ref{sec-nevsnh} below) or that $f$(H$_2$) is not a very reliable indicator of $n_{\rm H}$ or that the assumption of photoionization equilibrium does not hold.
The apparent lack of dependence on $f$(H$_2$) is particularly interesting for sightlines with relatively high molecular fraction (log[$f$(H$_2$)] $\ga$ $-$1.5) --- where we expect that the H$_2$ lines will be strong and self-shielded and where we do see variations in depletion.

\subsubsection{Is Ca I Enhanced Relative to Other Trace Neutrals?}
\label{sec-enhance}

In the main clouds toward 23 Ori, the electron density inferred from the ratio $N$(\ion{Ca}{1})/$N$(\ion{Ca}{2}) is much higher than the values derived from other such ratios (Welty et al. 1999b; see below) --- perhaps suggesting that some as yet undetermined process acts to enhance \ion{Ca}{1}, relative to the other trace neutral species.
In this section, we examine this issue for a larger sample of sightlines.

In photoionization equilibrium, we have 
\begin{equation}
\frac{N({\rm Ca~I})}{N({\rm X~I})} \sim 
\frac{(\Gamma/\alpha_r)_{\rm X}}{(\Gamma/\alpha_r)_{\rm Ca}}~
\frac{1}{\delta({\rm X}) {\rm A}_{\odot}({\rm X})}~
\frac{N({\rm Ca~II})}{N({\rm H})}.
\end{equation}
For the elements X = C, Na, K, and S, we would expect essentially linear relationships between $N$(\ion{Ca}{1})/$N$(\ion{X}{1}) and $N$(\ion{Ca}{2})/$N$(H), with only weak dependences on the temperature or radiation field, as the depletions of those four elements are thought to be uniform (within factors of 2).
In Figure~\ref{fig:xvsca}, we plot log[$N$(\ion{Ca}{1})/$N$(\ion{X}{1})] versus log[$N$(\ion{Ca}{2})/$N$(H)] for those four elements.
The solid lines of unit slope correspond to the relationships expected for solar (undepleted) abundances; the dotted lines correspond to the relationships expected for typical depletions of the four elements.
If grain-assisted recombination were much more important than radiative recombination for both calcium and the other four elements, the solid and dotted lines would shift vertically by less than about 0.2 dex in all cases.
If radiative recombination dominates for calcium but grain-assisted recombination dominates for the other four elements, then we would have
\begin{equation}
\frac{N({\rm Ca~I})}{N({\rm X~I})} \sim 
\frac{(\Gamma/\alpha_g)_{\rm X}}{(\Gamma/\alpha_r)_{\rm Ca}}~
\frac{n_e}{n_{\rm H}}~
\frac{1}{\delta({\rm X}) {\rm A}_{\odot}({\rm X})}~
\frac{N({\rm Ca~II})}{N({\rm H})}.
\end{equation}

While the correlation probabilities collected in Table~\ref{tab:corr} suggest that correlations are present in all four cases, the slopes of the relationships are rather uncertain. 
The regression fits to the detected points (weighted; denoted by an R in the last column of Table~\ref{tab:fits}) yield slopes ranging from 1.45 to 2.25, but the fits which attempt to account also for the limits (unweighted; denoted by an A)\footnotemark\ yield slopes from 0.5 to 0.9.
\footnotetext{We used the EM regression algorithm in ASURV, after removing any lower limits; the Buckley-James method yielded similar results.}
For comparison with the ``expected'' linear relationships, the dashed lines in Figure~\ref{fig:xvsca} show the fits to the detected points, with the slopes fixed at 1.0.
The observed data generally lie above the expected (depleted) lines, with mean offsets of factors of 3.2--6.9 --- as if $N$(\ion{Ca}{1}) were systematically higher than expected.
Because calcium is generally more severely depleted in the higher $N$(H) components (where the trace neutral species are more abundant), the $N$(\ion{Ca}{2})/$N$(H) ratio is likely to be lower there than for the line of sight as a whole --- making the discrepancies between the actual and predicted $N$(\ion{Ca}{1})/$N$(\ion{X}{1}) ratios even larger.
In all four cases, the Sco-Oph sightlines stand apart, with $N$(\ion{Ca}{1})/$N$(\ion{X}{1}) higher than for other sightlines at comparable $N$(\ion{Ca}{2})/$N$(H) [or $N$(\ion{Ca}{2})/$N$(H) lower for a given $N$(\ion{Ca}{1})/$N$(\ion{X}{1})].
That apparent regional difference and the overall considerable scatter --- ranges in $N$(\ion{Ca}{1})/$N$(\ion{X}{1}) of an order of magnitude or so at any given $N$(\ion{Ca}{2})/$N$(H) --- both suggest that there are real variations in the ratios that may reflect differences in environment (e.g., through differences in the relative contributions of radiative and grain-assisted recombination).

It is difficult to understand why \ion{Ca}{1} should be enhanced, relative to other trace neutral species, in many diffuse clouds.
As noted above, the generally small \ion{Ca}{1} $b$-values imply that any selective enhancement of \ion{Ca}{1} cannot be due to dielectronic recombination in warm ($T$ $\ga$ 3000 K) gas.
Dissociative recombination of CaH$^+$ does not appear to be likely, either (Weingartner \& Draine 2001).
If the ratio $\Gamma$/$\alpha_r$ has been overestimated for Ca$^0$ (by perhaps a factor of 4--5), then the predicted lines would lie below the data points in Figure~\ref{fig:xvsca}, as observed.
A spuriously high $\Gamma$/$\alpha_r$ would also yield systematically high electron densities from $N$(\ion{Ca}{1})/$N$(\ion{Ca}{2}), as found for a number of sightlines (see next section).
As already noted above (\S~\ref{sec-ionbal}), a much smaller true $\Gamma$/$\alpha_r$ would have potentially significant implications for both the ionization balance and the depletion of calcium where the $N$(\ion{Ca}{1})/$N$(\ion{Ca}{2}) ratio is small.
There are, however, some sightlines where \ion{Ca}{1} does not appear to be enhanced (e.g., $\zeta$ Oph; see below).
Some of the scatter (larger than expected from the nominal uncertainties) and possible regional variations seen in Figure~\ref{fig:xvsca} might be a result of differences in far-UV extinction and/or radiation field (e.g., for some of the Sco-Oph sightlines), given the differences in ionization potential and cross-section among the various neutral species (e.g., Jenkins \& Shaya 1979; Roberge et al. 1981; Welty 1989; WH).
It seems likely, however, that several (as yet undetermined) processes may contribute both to the systematic offsets and to the range of observed variations seen in Figure~\ref{fig:xvsca}.

\subsection{Electron Density and Fractional Ionization}
\label{sec-ne}

\subsubsection{Relative $n_e$ from Different X I/X II}
\label{sec-relne}

Several recent studies have suggested that the column density ratios $N$(\ion{X}{1})/$N$(\ion{X}{2}), for different elements X but for the same assumed temperature and radiation field, can yield markedly different values for $n_e$.
For various components toward the halo star HD 215733, Fitzpatrick \& Spitzer (1997) found systematic offsets between the electron densities derived from carbon (+0.4 dex) and magnesium (+0.15 dex; using more recent $f$-values for the \ion{Mg}{2} $\lambda$1239 doublet), relative to the values obtained from sulfur.
For the highest column density clouds toward 23 Ori, Welty et al. (1999b) found that the average $n_e$ derived independently from nine pairs of neutral and singly ionized species ranged from 0.04--0.05 cm$^{-3}$ (S, Mn, Fe) to 0.09--0.26 cm$^{-3}$ (C, Na, Mg, Si, K) to 0.95 cm$^{-3}$ (Ca).
After considering various possible explanations for those differences, Welty et al. conjectured that charge exchange between protons and the trace neutral species might account for the lower $n_e$ derived from sulfur, manganese, and iron, based on rather crude estimates for the relevant rates given by P\'{e}quignot \& Aldrovandi (1986).
No plausible explanation was found, however, for the high $n_e$ inferred from $N$(\ion{Ca}{1})/$N$(\ion{Ca}{2}).
Similar differences have been found for a few elements in the main absorption components toward the more heavily reddened stars HD 192639 and HD 185418 (Sonnentrucker et al. 2002 and in prep.).

In view of those observational results, Weingartner \& Draine (2001) have considered whether charge exchange between singly ionized atoms and neutral and/or negatively charged small grains (or large molecules such as polycyclic aromatic hydrocarbons) might play a significant role in the ionization balance of various elements (e.g., Lepp et al. 1988).
For plausible abundances of the small grains and models for the grain charging, Weingartner \& Draine find that this ``grain-assisted recombination'' can be comparable to radiative recombination in cold, diffuse clouds.
The corresponding recombination coefficients ($\alpha_g$) per hydrogen atom for such clouds, listed in the next-to-last column of Table~\ref{tab:atom}, range from 0.75--2.34 $\times$ 10$^{-14}$ cm$^3$ s$^{-1}$.
While that small range in $\alpha_g$ would seem to suggest that this process is no more element-specific than radiative recombination (compare the values of $\alpha_r$ in column 6 of Table~\ref{tab:atom}), significant differences can arise if different elements are characterized by very different sticking parameters ($s$).
This additional source of neutral species does reduce the electron densities inferred from some of the $N$(\ion{X}{1})/$N$(\ion{X}{2}) ratios, but it does not seem able to reconcile all the element-to-element differences in $n_e$.
In order to obtain more consistent $n_e$, Weingartner \& Draine argue that dissociative recombination of CH$^+$ may enhance \ion{C}{1} and that sodium and potassium could be essentially undepleted.\footnotemark\
With the additional assumption that $s$ = 0 for all elements, they thus find $n_e$ $\sim$ 0.03 cm$^{-3}$ for carbon, sodium, potassium, manganese, and iron in the main clouds toward 23 Ori, consistent with limits derived from the fine-structure excitation equilibrium of \ion{C}{2} and \ion{Si}{2}.
Significant, unexplained differences remain, however, for the (higher) electron densities derived from magnesium, silicon, and (especially) calcium.
\footnotetext{Welty et al. (1999b) had assumed depletions of $-$0.5 dex for both sodium and potassium --- slightly less severe than the values derived by WH from comparisons of \ion{C}{1}, \ion{Na}{1}, and \ion{K}{1}, using the observed (apparently constant) depletion of carbon and the assumption of photoionization equilibrium.
If grain-assisted recombination (with rates $\alpha_g$) were to dominate over radiative recombination, then those same comparisons would imply that both sodium and potassium typically are depleted by about $-$0.45 dex.
If, however, the more stringent criterion for electron transfer sketched by Weingartner \& Draine is adopted, then the depletions of both sodium and potassium could be even less severe, as the modified rates ($\alpha^{'}_{g}$) for those two elements appear to be more significantly reduced in cold, neutral clouds (see their Fig. 2).
In principle, other possible processes, such as dissociative recombination of CH$^+$, would also affect these depletion estimates.}

Because a number of the trace neutral species are significantly stronger than usual (relative to hydrogen) toward 23 Ori, one might ask whether the trends in $n_e$ found in that line of sight are representative of interstellar clouds in general.
In Table~\ref{tab:relne}, we list column densities for a number of neutral and singly ionized species, based on high-resolution optical and UV spectra, for the main interstellar components toward 23 Ori and four other stars with fairly extensive data for the trace neutral species.
The UV spectra --- generally obtained with the {\it HST}/GHRS --- were retrieved from the STScI archive and were fitted using profile structures determined from higher resolution optical spectra (as for 23 Ori).
Values in parentheses for some of the dominant first ions have been estimated from $N$(\ion{Zn}{2}) and typical depletions (Welty et al. 1999b; WH).
The electron densities inferred from the ratios $N$(\ion{X}{1})/$N$(\ion{X}{2}) are given on the third line in each set of values.
In each case, the first entry was derived assuming photoionization equilibrium at $T$ = 100 K, with the standard WJ1 radiation field (multiplied by 5.0 for 1 Sco and $\delta$ Sco; see \S~\ref{sec-fract}).
The left-hand panel of Figure~\ref{fig:relne} shows those $n_e$ [= $n_e$(rad)] for the five lines of sight, relative to the respective averages ($n_0$) of the values for carbon, sodium, and potassium [which are not entirely independent, as the typical depletions of sodium and potassium were estimated via the observed correlations between the corresponding trace neutral species (WH)].
In effect, the relative electron densities are given by
\begin{equation}
\frac{n_e}{n_0} \sim
\frac{(\Gamma/\alpha_r)_{\rm X}}{(\Gamma/\alpha_r)_{\rm CNaK}}
\frac{N({\rm X~I})}{N({\rm CNaK~I})}
\frac{N({\rm CNaK~II})}{N({\rm X~II})},
\end{equation}
where the values for ``CNaK'' represent suitable averages for carbon, sodium, and potassium; we have assumed (for the present) that radiative recombination dominates.
These relative $n_e$ should be fairly insensitive to the actual values of $T$ and the ambient radiation fields characterizing the clouds in the different lines of sight, as well as to differences in distribution between the neutral and singly ionized species (\S~\ref{sec-approx}; \S~\ref{sec-ca1k1}).

In order to gauge the effects of grain-assisted recombination on the inferred electron densities (for each element in each line of sight), we solved equation 16 of Weingartner \& Draine (2001) (given above, in \S~\ref{sec-approx}).
We have assumed that the depletions of sodium and potassium are less severe by about 0.2 dex (as would be the case if grain-assisted recombination dominates); values of $n_{\rm H}$ are given for these five lines of sight in Table~\ref{tab:pdfi}.
The second $n_e$ entries for each element in Table~\ref{tab:relne} give the values obtained assuming that both radiative and grain-assisted recombination must be considered; the right-hand panel of Figure~\ref{fig:relne} shows the corresponding relative $n_e$ (again normalized to the average of the values for carbon, sodium, and potassium).
The entries in the table and the circles in the figure correspond to our assumed sticking parameters:  $s$ = 0 for carbon, sodium, sulfur, and potassium (little depleted); $s$ = 0.5 for magnesium, silicon, and manganese (moderately depleted); and $s$ = 1 for calcium and iron (heavily depleted).
(Weingartner \& Draine did not consider lithium, aluminum, chromium, or nickel.)
The associated ``error bars'' (attached via vertical dotted lines) show the range in relative $n_e$ if we assume either $s$ = 0 or $s$ = 1 for all elements.
In principle, those two extremes give the maximum and minimum effect of grain-assisted recombination; the circles in Figure~\ref{fig:relne} do not always lie between the values for $s$ = 0 and 1 because we have assumed different values of $s$ for different elements and because we have normalized by the mean $n_e$ for carbon, sodium, and potassium (which all have $s$ = 0).
The most significant reductions in $n_e$ generally are seen for carbon, sodium, sulfur, and potassium (all with $s$ = 0); the less severe depletions for sodium and potassium also play a role.
For example, the electron densities inferred from those four elements for the main clouds toward $\zeta$ Oph are smaller by factors of 3--12 than the values obtained by assuming only radiative recombination; somewhat smaller effects are seen for the other four lines of sight.
For Ca$^0$, the effect of grain-assisted recombination is generally very minor, as the ``radiative'' $n_e$ are typically fairly large and (especially) as we have adopted $s$ = 1.

It is apparent from Figure~\ref{fig:relne} that the differences in relative $n_e$ found toward 23 Ori are neither unique nor universal.
The ``radiative'' electron densities inferred for the main clouds toward $\zeta$ Ori, for example, are significantly lower than those toward 23 Ori, but show a very similar pattern --- with roughly similar values from carbon, sodium, magnesium, and potassium (0.03--0.07 cm$^{-3}$), low values from sulfur and iron ($<$0.008 cm$^{-3}$), and a high value from calcium (0.30 cm$^{-3}$).
The patterns seen toward 1 Sco and $\delta$ Sco are also similar, except that the electron densities derived from silicon and iron are comparable to the values derived from carbon, sodium, magnesium, and potassium --- rather than much lower, as seen toward 23 Ori.
In the main clouds toward $\zeta$ Oph, however, most elements --- even sulfur and calcium --- give $n_e$ in the range 0.19--0.41 cm$^{-3}$, and only iron gives a low value (0.03 cm$^{-3}$).
While including grain-assisted recombination (right panel) gives slightly better agreement for silicon, sulfur, manganese, and iron toward 23 Ori, the agreement among the various elements toward $\zeta$ Oph is poorer (except that the value for iron is more consistent).
In general, the values for the depleted elements (with $s$ $>$ 0) are increased, relative to those for the lightly depleted elements (with $s$ $\sim$ 0).
Overall, the patterns of relative $n_e$ seem rather similar for the two cases.

If viewed by element instead of by sightline, the electron densities inferred from calcium are often higher than the values derived from other elements [but not toward $\zeta$ Oph, unless grain-assisted recombination is important (for other elements)].
The values obtained from iron (and perhaps manganese and nickel) are often low (but not toward 1 Sco or $\delta$ Sco); some apparent deficiencies of \ion{Mn}{1} and \ion{Fe}{1} had previously been reported by Joseph, Snow, \& Seab (1989) and Snow, Joseph, \& Meyer (1986).
The values derived from sulfur are also often low (but not toward $\zeta$ Oph).
Unfortunately, there are currently only a few sightlines with reliable high-resolution data for both neutral and singly ionized species for many elements.

The differences in the element-to-element pattern of relative $n_e$ for these five lines of sight suggest that the differences in electron density inferred for any given line of sight are not just due to a single process or factor --- but instead result from multiple additional processes that selectively affect the ionization balance of the various elements, to different degrees in the different lines of sight.
The similarities between the patterns seen for (23 Ori, $\zeta$ Ori), and also for (1 Sco, $\delta$ Sco) --- and the differences between the Orion, Scorpius, and $\zeta$ Oph patterns --- are suggestive of regional variations in some (as yet undetermined) physical parameter(s).
Fitzpatrick \& Spitzer (1997), Welty et al. (1999b), and Weingartner \& Draine (2001) have all described some possible factors and/or processes --- including problems with the adopted rates, dielectronic recombination, dissociative recombination, density-dependent depletions, charge exchange with small grains, and charge exchange with protons --- but consistency among the various estimates for $n_e$ remains elusive.

\subsubsection{Dissociative Recombination of CH$^+$ as a Source of C I}
\label{sec-dissrec}

In principle, dissociative recombination of CH$^+$ ``should'' have a measurable impact on the ionization balance of carbon.
The dissociative recombination rate has been estimated as $\alpha_d$ $\sim$ 3.3 $\times$ 10$^{-7}$ cm$^3$ s$^{-1}$ at $T$ $\sim$ 100 K (Mitchell \& McGowan 1978).
Since $N$(\ion{C}{2})/$N$(H) $\sim$ 1.4 $\times$ 10$^{-4}$ and $N$(CH$^+$)/$N$(H) ranges from about 1--25 $\times$ 10$^{-9}$ (from examination of various references), the ratio [$\alpha_d$ $N$(CH$^+$)]/[$\alpha_r$ $N$(\ion{C}{2})] --- which gives the relative contributions of dissociative and radiative recombination to \ion{C}{1} --- could be between 0.3 and 6.6.
If dissociative recombination were generally dominant, we would expect that $N$(\ion{C}{1})/$N$(\ion{X}{1}) would be proportional to $N$(CH$^+$)/$N$(H), for some element X whose depletion is roughly constant.
In Figure~\ref{fig:dissrec}, we show the relationships between those ratios, for X = K, Na, and S. 
Of those ratios, consideration of the ionization potentials of the neutral species suggests that $N$(\ion{C}{1})/$N$(\ion{S}{1}) should be the least susceptible to any effects of differences in the far-UV extinction.
While the correlation probabilities (Table~\ref{tab:corr}) indicate that there is no correlation for \ion{S}{1} or \ion{K}{1}, the two methods which include limits suggest a possible correlation for \ion{Na}{1}.
The regression fit, however, yields a slope consistent with no correlation (Table~\ref{tab:fits}); including the available limits seems unlikely to increase it significantly.
Probable differences in distribution between CH$^+$ and hydrogen might act to obscure any correlations, however.
As noted by Welty et al. (1999b) for 23 Ori, CH$^+$ may not be entirely coextensive with the various trace neutral species, and so would not have as strong an impact on the ionization balance of carbon.
Inclusion of grain-assisted recombination will reduce the relative importance of dissociative recombination and (probably) add scatter in comparing the ratios considered in this section --- making it more difficult to discern the possible effects of dissociative recombination. 

Even if dissociative recombination of CH$^+$ is not the dominant source of \ion{C}{1}, however, it may play a significant role in some clouds.
Table~\ref{tab:relne} and Figure~\ref{fig:relne} both indicate that the electron density derived from carbon may be slightly higher than the values obtained from sodium and potassium for 23 Ori and $\zeta$ Oph [where $N$(CH$^+$) is greater than 10$^{13}$ cm$^{-2}$], but is slightly lower for $\zeta$ Ori, 1 Sco, and $\delta$ Sco [where $N$(CH$^+$) is $\le$ 3 $\times$ 10$^{12}$ cm$^{-2}$].
[The shallow far-UV extinction toward $\delta$ Sco and other stars in that region may also contribute to lower $N$(\ion{C}{1}), however (Welty 1989).]
Enhanced \ion{C}{1} toward some stars in the Pleiades (WH) may also be related to the relatively high $N$(CH$^+$) there (White 1984; Crane, Lambert, \& Sheffer 1995).

\subsubsection{Does $n_e$ Depend on $n_{\rm H}$?}
\label{sec-nevsnh}

Different sources for the electrons in diffuse interstellar clouds have corresponding different dependences of $n_e$ on the local hydrogen density $n_{\rm H}$.
Those dependences, in turn, would yield different relationships between the local densities (and corresponding column densities) of various trace neutral species and those of hydrogen.  
Table~\ref{tab:pdfi}, lists local hydrogen densities, electron densities, and fractional ionizations ($n_e$/$n_{\rm H}$) estimated for the main clouds toward twelve stars; values for other discernible component groups are also given for several of the sightlines.
The estimates for $n_{\rm H}$ are obtained from thermal pressures ($n_{\rm H}T$) derived from analyses of the \ion{C}{1} fine-structure excitation equilibrium (Jenkins \& Shaya 1979). 
The \ion{C}{1} column densities are derived from high-resolution (FWHM $\sim$ 3 km~s$^{-1}$) UV spectra obtained with {\it HST}, in order to obtain accurate values for specific component groups (Wannier et al. 1999; Welty et al. 1999b and unpublished; Zsarg\'{o} 2000; Jenkins \& Tripp 2001; Sonnentrucker et al. 2002 and in prep.).
The corresponding temperatures are obtained from analyses of the rotational excitation of H$_2$ (Savage et al. 1977; Welty et al. 1999b; Rachford et al. 2002); values in parentheses are somewhat arbitrarily assumed (as 100 K).
The last four columns give the observed $N$(\ion{Ca}{1})/$N$(\ion{Ca}{2}) ratio, then two sets of values for the electron density inferred from that ratio, the average electron density derived from the corresponding ratios for carbon, sodium, and potassium (assuming typical depletions where necessary), and the fractional ionization using that average $n_e$.
The first set assumes only radiative recombination; the second set includes both radiative and grain-assisted recombination, as described above (\S\S~\ref{sec-approx} and \ref{sec-relne}).
Electron densities in parentheses have been estimated from total sightline column densities --- and so should be viewed as lower limits to the values in the gas traced by the neutral species.
For all the electron density estimates, we use radiative recombination coefficients ($\alpha_r$) corresponding to the adopted temperature. 
We assume the average (WJ1) interstellar radiation field for all cases except the three Scorpius sightlines, where we multiply that field by a factor 5 (one possible explanation for the lower than usual abundances of some trace neutral species seen in those sightlines; see WH).
The assumption of the unattenuated field should give at least rough estimates for $n_e$ in these relatively diffuse clouds [with individual component $E(B-V)$ $<$ 0.2].

The values for $n_{\rm H}$, $n_e$, and $n_e$/$n_{\rm H}$ collected in Table~\ref{tab:pdfi} reveal several interesting trends.
In the upper panel of Figure~\ref{fig:nevsnh}, we plot the values of $n_e$ determined from calcium (squares) and from carbon, sodium, and potassium (CNaK) --- with and without grain-assisted recombination (circles and triangles, respectively) ---  versus the local density $n_{\rm H}$.
The open symbols denote the three sightlines in Scorpius, where the radiation field was assumed to be 5$\times$ stronger.
For this relatively small sample, the density ranges from 6--335 cm$^{-3}$ (a factor greater than 50), but $n_e$(rad) (CNaK) ranges only from 0.04--0.23 cm$^{-3}$ (a factor of 6) --- with no apparent dependence of $n_e$(rad) on $n_{\rm H}$.
The electron densities computed when grain-assisted recombination is included, however, appear to be somewhat smaller for larger $n_{\rm H}$ --- reflecting the increased effect of charge exchange at higher densities.
On the other hand, the $n_e$(rad) derived from calcium (which is not significantly affected by grain-assisted recombination, if the sticking factor $s$ = 1) may show slight increases with increasing density --- but this result depends somewhat on the assumed stronger radiation fields for the Scorpius sightlines.
We note, however, that the ratio $n_e$(Ca)/$n_e$(CNaK) is also higher for those three sightlines (relatively independent of the strength of the radiation field).
Similar trends are seen versus the thermal pressure $n_{\rm H}T$, as the temperature ranges only over a factor of 2 for this small sample.
While the sample of component groups in Table~\ref{tab:pdfi} is likely not entirely representative of the diffuse ISM [a number of the groups have thermal pressures much higher than the typical values (Jenkins \& Tripp 2001)], these trends appear to be present even when the higher pressure groups are excluded.

The apparent lack of dependence of $n_e$(rad) on $n_{\rm H}$ and the possible inverse relationship between $n_e$ and $n_{\rm H}$ when grain-assisted recombination is included are rather surprising.
If the electrons are due primarily to photoionization of carbon and other abundant heavy elements, we would expect $n_e$ $\propto$ $n_{\rm H}$.
If the electrons instead are due primarily to cosmic ray ionization of hydrogen, we would expect $n_e$ $\propto$ $n_{\rm H}^{1/2}$ (e.g., Hobbs 1974; Kulkarni \& Heiles 1987).
From the discussions above (\S\S~\ref{sec-approx} and ~\ref{sec-relne}), we also would expect $n_e$(rad) = $n_e$ + ($\alpha_g$ $n_{\rm H}$)/$\alpha_r$ for $s$ = 0 (as adopted for carbon, sodium, and potassium), with the second term often dominant.
We note, however, that some of the models of Liszt (2003) --- particularly those with enhanced cosmic ray ionization; mild depletion of carbon, nitrogen, and oxygen; and less than the nominal grain-assisted recombination --- predict relatively constant $n_e$ for $n_{\rm H}$ $\la$ 100 cm$^{-3}$.

A lack of dependence of $n_e$ on $n_{\rm H}$ would also make it difficult to understand some of the observed relationships between the column densities of \ion{X}{1}, \ion{X}{2}, and H, for the typically severely depleted elements X = Fe and Ca.
If radiative recombination dominates for iron and calcium (with $s$ $\sim$ 1), then we would expect that $n$(X$^0$) $\propto$ $n_e$ $n$(X$^+$) $\propto$ $n_e$ $n_{\rm H}$ $\delta$(X).
The relationships reported by Wakker \& Mathis (2000) suggest that $N$(\ion{Fe}{2}) $\propto$ [$N$(H)]$^{0.4}$ and $N$(\ion{Ca}{2}) $\propto$ [$N$(H)]$^{0.2}$ --- presumably reflecting increasingly severe depletions in higher column density (denser?) gas.
Combining column densities for \ion{Fe}{1} and \ion{Ca}{1} from this paper with $N$(H) from WH, we obtain $N$(\ion{Fe}{1}) $\propto$ [$N$(H)]$^{1.37\pm0.20}$ and $N$(\ion{Ca}{1}) $\propto$ [$N$(H)]$^{1.25\pm0.14}$.
Taking the ratios of the neutral and singly ionized species, we thus find that $N$(\ion{X}{1})/$N$(\ion{X}{2}) --- usually assumed to be proportional to $n_e$ --- is roughly proportional to $N$(H) for both calcium and iron.

Hobbs (1974) had noted that the observed roughly quadratic dependences of $N$(\ion{K}{1}) and $N$(\ion{Na}{1}) on $N$(H) would be consistent with the essentially constant fractional ionization expected if most electrons came from photoionization of carbon, and concluded that cosmic ray ionization of hydrogen [which would yield shallower relationships between those $N$(\ion{X}{1}) and $N$(H) under the assumption of photoionization equilibrium] must not be the main source of electrons.
If grain-assisted recombination dominates the production of the trace neutral species for very mildly depleted elements like sodium and potassium (with $s$ $\sim$ 0), however, then we would expect that $n$(X$^0$) $\propto$ $n_{\rm H}$ $n$(X$^+$) $\propto$ $n_{\rm H}^2$ $\delta$(X) --- consistent with the observed roughly quadratic relationships.
Those observed quadratic relationships thus would not exclude significant cosmic ray ionization of hydrogen in diffuse interstellar clouds (Liszt 2003).

\subsubsection{Fractional Ionization}
\label{sec-fract}

The other puzzling aspect of the ionization balance in the main clouds toward 23 Ori is the generally high fractional ionization --- of order 0.01 --- implied by the adopted $n_e$ and the $n_{\rm H}$ obtained from analysis of the \ion{C}{1} fine structure excitation.
That fractional ionization is a factor 50 higher than the value $\sim$ 2 $\times$ 10$^{-4}$ expected from photoionization of carbon and other abundant heavy elements, and is still a factor of order 10 higher than the values obtained by including ionization of hydrogen and helium due to cosmic rays and x-rays, even if that ionization rate is as high as 10$^{-16}$ s$^{-1}$ (e.g., Kulkarni \& Heiles 1987; Wolfire et al. 1995; Weingartner \& Draine 2001).
Welty et al. (1999b) had conjectured that charge exchange with large molecules might have enhanced the trace neutral species by perhaps an order of magnitude (based on the models of Lepp et al. 1988), but the more detailed calculations of Weingartner \& Draine (2001) suggest that the effects of charge exchange will be somewhat smaller (at least in the main clouds toward 23 Ori; see Table~\ref{tab:relne}).
In addition, Weingartner \& Draine note that grain-assisted recombination will also affect the abundance of H$^+$ in predominantly neutral regions --- countering the effects of cosmic ray and x-ray ionization of hydrogen to some extent (see also Liszt 2003).
The question (again) is how representative the clouds toward 23 Ori are of the diffuse ISM.

In the lower panel of Figure~\ref{fig:nevsnh}, we plot the fractional ionizations for the component groups listed in Table~\ref{tab:pdfi} versus $n_{\rm H}$, using the same symbols as in the upper panel. 
While $n_e$/$n_{\rm H}$ is often assumed to be constant, the values for the small sample shown in the figure appear to decrease at higher densities (corresponding to the trends seen for $n_e$ in the upper panel).
Nine of the 13 component groups have $n_e$(rad)/$n_{\rm H}$ $\ga$ 10$^{-3}$ (triangles); three of the groups (those toward HD 185418, $\zeta$ Per, and 23 Ori) exhibit values greater than 10$^{-2}$.
Inclusion of grain-assisted recombination (circles) reduces the values for three of those nine component groups to within the range 2--10 $\times$ 10$^{-4}$ ``expected'' if both photoionization of carbon (and other abundant heavy elements) and cosmic ray ionization of hydrogen are significant sources of electrons, but also reduces the other four values originally within that range to well below 2 $\times$ 10$^{-4}$.

\subsubsection{Electron Densities from $N$(\ion{Ca}{2})/$N$(\ion{Ti}{2})}
\label{sec-stokes}

In connection with a large survey of interstellar \ion{Ti}{2} absorption, Stokes (1978) developed a method for estimating $n_e$ from the ratio $N$(\ion{Ca}{2})/$N$(\ion{Ti}{2}) (see also Stokes \& Hobbs 1976).
This method would be of particular utility for the thinner diffuse clouds, where \ion{Ca}{2} is likely a trace species (with \ion{Ca}{3} dominant) and where many of the various trace neutral species are more difficult to detect.
The method assumes both photoionization equilibrium and a constant ratio $r$ = [$A_{\odot}$(Ca)$\delta$(Ca)]/[$A_{\odot}$(Ti)$\delta$(Ti)] in the interstellar gas [as suggested by the good correlation observed between $N$(\ion{Ca}{2}) and $N$(\ion{Ti}{2})].
The value of $r$ is calibrated by the $N$(\ion{Ca}{1})/$N$(\ion{Ca}{2}) ratio (or the corresponding $n_e$) where \ion{Ca}{1} has been detected.
We then have $n_e$ $\sim$ ($\Gamma$/$\alpha_r$) \{$r$[$N$(\ion{Ti}{2})/$N$(\ion{Ca}{2})] $-$ 1\}$^{-1}$, where the $\Gamma$/$\alpha_r$ is for the \ion{Ca}{2}--\ion{Ca}{3} equilibrium.

The above discussions suggest several potential difficulties with this method, however.
First, the assumption of photoionization equilibrium may not hold, as other processes may be significant (though grain-assisted recombination may not be important for these typically severely depleted elements).
Second, calibrating $r$ with the $n_e$ from calcium will yield a value that is too small (and thus estimated electron densities that are too large).
Finally, in order to estimate $n_e$ (given $r$), one must know the temperature (in order to obtain the recombination rate $\alpha_r$ for Ca$^{++}$ + $n_e$).
If the temperature can be as high as several thousand K for warm, neutral gas (where much of the \ion{Ti}{2} may reside), then $\alpha_r$ could be as much as an order of magnitude smaller than the value for $T$ $\sim$ 100 K which is usually assumed.

\subsection{Density Dependence of Depletions}
\label{sec-cfs}

As noted above, trends in average line-of-sight depletions with average density $<n_{\rm H}>$ and with $f$(H$_2$) have suggested that the depletions may depend (to varying degrees) on the local hydrogen density $n_{\rm H}$.
Using column density data collected from the literature and/or derived from new observations of stars in the Cepheus OB3 association, Cardelli et al. (1991) found an inverse relationship between the ratios $N$(CN)/$N$(CH) and $N$(\ion{Ca}{1})/$N$(CH).
Cardelli et al. surmised that this somewhat unexpected anti-correlation between \ion{Ca}{1} and CN might be due to increasingly severe depletion of calcium in the denser gas responsible for the CN.
Using a simple model for the gas-phase chemistry and assuming that photoionization equilibrium holds for calcium and that most electrons come from ionization of carbon, they concluded that the calcium depletion varies roughly as $n_{\rm H}^{-3}$.
This method thus seems able to provide specific relationships between the local density and the depletions of various elements X, if column densities for \ion{X}{1}, CH, and CN are available for a sufficient sample of sightlines.
Again, however, the above discussions suggest that the assumptions of photoionization equilibrium and of carbon being the primary source of electrons may not be valid.

Using the larger sample of column densities now available, we may examine such relationships for several different trace neutral species (Figure~\ref{fig:cfs1}).
The correlation probabilities obtained from the survival analysis routines suggest that (anti)correlations may be present for \ion{Ca}{1} and \ion{Fe}{1} (Table~\ref{tab:corr}) --- though the values obtained from the detected points alone suggest otherwise.
The relationship between $N$(CN)/$N$(CH) and $N$(\ion{Ca}{1})/$N$(CH) seems both steeper (slope $\la$ $-$2.5 rather than $-$1; Table~\ref{tab:fits}) and less tight than the relationship plotted by Cardelli et al. (1991).\footnotemark\
\footnotetext{While Cardelli et al. (1991) considered upper limits essentially as detections, we have chosen to fit only the points corresponding to detections --- which may contribute to the steeper slope derived in this paper.
For all three trace neutral species considered here, a slope of infinity yields a better fit to the available data than the slope of $-$1 found for \ion{Ca}{1} by Cardelli et al.}
Within the Cardelli et al. framework, steeper relationships correspond to weaker dependences of the depletion on $n_{\rm H}$; a slope of $-$2.5 would imply that the calcium depletion is proportional to $n_{\rm H}^{-1.8}$.
Substituting \ion{Fe}{1} for \ion{Ca}{1} also yields a steeper slope ($\la$ $-$2.0), with a slightly weaker (anti)correlation than for \ion{Ca}{1}.
For $N$(CN)/$N$(CH) vs. $N$(\ion{K}{1})/$N$(CH), we find an essentially infinite slope [omitting the points for HD 37903 and HD 73882 with anomalously small $N$(\ion{K}{1})/$N$(CH)], since the latter ratio appears to be roughly constant for most lines of sight (WH).
The analog of equation 10 in Cardelli et al. for potassium would thus imply that the depletion of potassium is proportional to $n_{\rm H}^{-1}$.
The roughly constant ratio $N$(\ion{C}{1})/$N$(\ion{K}{1}) (WH) and the apparently uniform depletion of carbon over wide ranges in $<n_{\rm H}>$ and $f$(H$_2$), however, suggest that the depletion of potassium does not depend on density, at least for the range in density probed so far.\footnotemark\
\footnotetext{The possible upturn in $N$(\ion{C}{1}) vs. $N$(\ion{K}{1}) for the highest column densities in the current sample may reflect uncertainties in the $f$-values for some of the \ion{C}{1} lines (Jenkins \& Tripp 2001) and/or the effects of UV extinction (Jenkins \& Shaya 1979; Welty 1989).}

Another way of gauging the possible dependence of depletions on density is to compare ratios such as $N$(\ion{Ca}{1})/$N$(\ion{K}{1}) [$\propto$ $\delta$(Ca)] with the ratio $N$(CN)/$N$(CH).
Because relatively dense gas is necessary for the formation of CN, the ratio $N$(CN)/$N$(CH) should be a density indicator; simple chemical models suggest that $n$(CN)/$n$(CH) is proportional to $n_{\rm H}^2$ (Cardelli et al. 1991).
The plots of $N$(\ion{Ca}{1})/$N$(\ion{K}{1}) and $N$(\ion{Fe}{1})/$N$(\ion{K}{1}) vs. $N$(CN)/$N$(CH) shown in Figure~\ref{fig:cfs2} reveal (at best) relatively weak anti-correlations (Table~\ref{tab:corr}); only the correlation tests which include limits suggest that correlations may be present.
The slopes computed for the detected points are marginally less than zero (Table~\ref{tab:fits}); the available limits are consistent with those fits. 
Similar behavior is observed if \ion{Na}{1} is substituted for \ion{K}{1} (but for fewer lines of sight).
If $n$(CN)/$n$(CH) goes as $n_{\rm H}^{2}$ and if the depletions of sodium and potassium do not depend on the density, then these weak anti-correlations would be consistent with more severe depletion of calcium and iron in denser gas --- but with a much weaker dependence on density (at most as $n_{\rm H}^{-0.5}$) than that found for calcium by Cardelli et al.

Finally, for the small sample of component groups listed in Table~\ref{tab:pdfi}, we may directly compare $N$(\ion{Ca}{2})/$N$(H) with $n_{\rm H}$.
In all cases, the $N$(\ion{Ca}{1})/$N$(\ion{Ca}{2}) ratios suggest that \ion{Ca}{2} is the dominant form of calcium in those component groups, so that $N$(\ion{Ca}{2})/$N$(H) $\propto$ $\delta$(Ca).
A weighted fit to the 11 data points yields the relation $\delta$(Ca) $\propto$ [$n_{\rm H}$]$^{-1.0\pm0.4}$; any differences in distribution between \ion{Ca}{2} and H would likely make the dependence somewhat shallower.

\subsection{Individual Lines of Sight}
\label{sec-indiv}

\subsubsection{Cepheus OB2}
\label{sec-cepob2}

Many of the strongest \ion{Ca}{1} lines observed in this survey are toward stars in the Cepheus OB2 association.
In several of the Cep OB2 sightlines --- toward HD 207198, $\nu$ Cep, and $\lambda$ Cep --- there is a strong, narrow \ion{Ca}{1} component, with relatively high $N$(\ion{Ca}{1})/$N$(\ion{K}{1}), at or near the red-ward edge of the absorption (though at somewhat different velocities in the three cases).
The red-most \ion{Ca}{1} component toward $\lambda$ Cep, at $v$ $\sim$ $-$2.7 km~s$^{-1}$, also has one of the highest $N$(\ion{Ca}{1})/$N$(\ion{Ca}{2}) ratios in the sample.
Analysis of the \ion{C}{1} fine-structure excitation toward $\lambda$ Cep suggests that the $-$2.7 km~s$^{-1}$ component is characterized by a somewhat greater thermal pressure than the other strong, low-velocity components, which have log $n_{\rm H}T$ (cm$^{-3}$K) of order 3.3 (Table~\ref{tab:pdfi}; Jenkins \& Tripp 2001).

Jenkins \& Tripp (2001) find even higher pressures for the component near $-$35 km~s$^{-1}$ toward $\lambda$ Cep.
The \ion{C}{1} analysis yields pressures ranging from log $n_{\rm H}T$ $\sim$ 5.3 for $T$ = 160 K to log $n_{\rm H}T$ $\sim$ 5.8 for $T$ = 80 K; similar analysis of the \ion{O}{1} fine-structure excitation yields log $n_{\rm H}T$ $\sim$ 4.8.
The relatively narrow width of that component as seen in \ion{K}{1} --- $b$ $\sim$ 0.8 km~s$^{-1}$ (WH) --- implies that the temperature must be less than about 1500 K; if the turbulence in the gas is just sonic, then $T$ would be of order 150 K.
Intriguingly, the $N$(\ion{Ca}{1})/$N$(\ion{Ca}{2}) ratio --- and thus $n_e$/($\Gamma$/$\alpha_r$)? --- in the $-$35 km~s$^{-1}$ component is not much greater than the values found for the low-velocity components --- even though the pressure is at least 30 times higher.

\subsubsection{Orion}
\label{sec-orion}

The narrow blue-most \ion{Ca}{1} components toward $\iota$ Ori, $\epsilon$ Ori, and $\zeta$ Ori (at slightly different velocities) are quite strong, relative to the corresponding narrow components in \ion{Na}{1} and \ion{K}{1} (WHK: WH; Price et al. 2001).
With \ion{Na}{1} $b$-values of order 0.35--0.45 km~s$^{-1}$, these components must have $T$ $\la$ 275 K.
Toward $\epsilon$ Ori, the component at 3.1 km~s$^{-1}$ has a high $N$(\ion{Ca}{1})/$N$(\ion{Ca}{2}) ratio --- nearly 8 times higher than the value for the main components near 24 km~s$^{-1}$.
Toward $\zeta$ Ori, however, the component at $-$1.8 km~s$^{-1}$ has a somewhat lower $N$(\ion{Ca}{1})/$N$(\ion{Ca}{2}) ratio, more comparable to that in the main components near 24 km~s$^{-1}$.
We note, however, that this component is apparently time-variable (Price et al. 2001), and that the corresponding $N$(\ion{Ca}{2}) may actually be significantly smaller than the value used here, as the \ion{Ca}{2} component at that velocity (in the published profile fits) has a much higher $b$-value (WMH; Price et al. 2001).
Analysis of the \ion{C}{1} fine-structure excitation toward $\zeta$ Ori indicates that the thermal pressure in the $-$1.8 km~s$^{-1}$ component is more than an order of magnitude larger than that in the main components (Table~\ref{tab:pdfi}).
Jenkins \& Peimbert (1996) found a systematic broadening, with increasing rotational level $J$, of the H$_2$ lines near $v$ $\sim$ 0 km~s$^{-1}$ in IMAPS spectra of $\zeta$ Ori, and proposed that the H$_2$ is formed in cooling, compressed post-shock gas (see also Welty et al. 2002).
The gas-phase abundances (relative to S and Zn) in the gas near $v$ = 0 km~s$^{-1}$ are very similar to those typically found in warm, quiescent, diffuse clouds, however --- i.e., there are no obvious enhancements of typically severely depleted elements that would suggest that that gas has been shocked.

\subsubsection{SN 1987A}
\label{sec-sn87a}

Magain (1987) found unusually strong \ion{Ca}{1} absorption in a number of velocity components toward the LMC SN 1987A. 
In particular, the ratios $N$(\ion{Ca}{1})/$N$(\ion{Na}{1}) and $N$(\ion{Ca}{1})/$N$(\ion{Mg}{1}) in some of the component groups are significantly higher than the values commonly found in the Galactic ISM; the $N$(\ion{Ca}{1})/$N$(\ion{Ca}{2}) ratio, however, generally is closer to the Galactic values (Welty et al. 1999a).
While much of the apparent enhancement of \ion{Ca}{1} toward SN 1987A is likely a result of less severe calcium depletion, dielectronic recombination of \ion{Ca}{2} in relatively warm gas ($T$ $\sim$ 4500 K) may also be a factor for some of the intermediate-velocity components (Welty et al. 1999a).

\section{Summary}
\label{sec-sum}

We have analyzed high-resolution (FWHM $\sim$ 0.3--1.5 km~s$^{-1}$) spectra of interstellar \ion{Ca}{1} $\lambda$4226 absorption toward 30 Galactic stars.
Typical 2$\sigma$ equivalent width limits for narrow individual components are 0.16--0.95 m\AA, which correspond to \ion{Ca}{1} column density limits of 0.6--3.6 $\times$ 10$^9$ cm$^{-2}$.
These spectra are thus characterized by higher resolution and sensitivity than the \ion{Ca}{1} spectra previously available for most of these lines of sight.

Voigt-profile fits to the observed line profiles yielded a total of 112 individual \ion{Ca}{1} components --- which generally correspond to components seen in \ion{Ca}{2}, \ion{Na}{1}, and \ion{K}{1}.
The median \ion{Ca}{1} $b$-value for the highest resolution spectra, 0.66 km~s$^{-1}$, is comparable to those found for \ion{Na}{1} (WHK) and \ion{K}{1} (WH), as would be expected if the three species are essentially coextensive.
The generally larger $b$-values found for \ion{Ca}{2} (WMH) suggest that \ion{Ca}{2} may be more broadly distributed, in gas characterized by somewhat higher temperatures and/or turbulent velocities.
For corresponding individual components at the same velocity, the ratio $N$(\ion{Ca}{1})/$N$(\ion{K}{1}), indicative of the calcium depletion, ranges over a factor of about 500; the ratio $N$(\ion{Ca}{1})/$N$(\ion{Ca}{2}), equal to $n_e$/($\Gamma$/$\alpha_r$) in photoionization equilibrium, ranges over a factor of about 50.
Comparisons between \ion{Ca}{1} and various other species yield the following results:

\begin{itemize}

\item{If photoionization and radiative recombination dominate the calcium ionization balance, then the observed column densities of \ion{Ca}{1} and \ion{Ca}{2} suggest that \ion{Ca}{2} is the dominant form of calcium wherever \ion{Ca}{1} is detected.
\ion{Ca}{3} could still be dominant, where $N$(\ion{Ca}{1})/$N$(\ion{Ca}{2}) is smallest, if charge exchange with small grains is more important than radiative recombination.}

\item{The ratio $N$(\ion{Ca}{1})/$N$(\ion{Ca}{2}) does not seem to depend on the fraction of hydrogen in molecular form $f$(H$_2$) --- suggesting either that $n_e$/($\Gamma$/$\alpha_r$) does not depend on $n_{\rm H}$ or that $f$(H$_2$) is not a reliable indicator of density or that photoionization equilibrium does not apply.}

\item{\ion{Ca}{1} seems to be systematically more abundant, relative to other trace neutral species, than would be expected if photoionization and radiative recombination control the ionization balance of heavy elements.
The typically small \ion{Ca}{1} $b$-values imply that the temperature cannot be high enough ($\ga$ 3000 K), in general, for dielectronic recombination of Ca$^{+}$ to account for that apparent overabundance.
Grain-assisted recombination is unlikely to significantly affect the ionization balance of the typically severely depleted calcium.}

\item{The electron densities inferred from the ratios $N$(\ion{X}{1})/$N$(\ion{X}{2}), for different elements X, can differ by more than an order of magnitude along a given line of sight, if photoionization equilibrium is assumed.
In particular, the $n_e$ obtained from $N$(\ion{Ca}{1})/$N$(\ion{Ca}{2}) is often significantly higher than the values derived from other elements.
The values obtained from the corresponding ratios for sulfur, manganese, iron, and nickel are often lower than those derived from carbon, sodium, and potassium.
Those differences are not eliminated if charge exchange with small grains dominates the production of (at least) some of the trace neutral species \ion{X}{1}.
The patterns of relative $n_e$ for different lines of sight show both similarities and differences --- suggesting that multiple additional processes besides photoionization and radiative recombination significantly affect the ionization balance of heavy elements --- to different degrees in different lines of sight.}

\item{The lack of a strong correlation between $N$(\ion{C}{1})/$N$(\ion{X}{1}) and $N$(CH$^+$)/$N$(H), for X = Na, K, and S, suggests that dissociative recombination of CH$^+$ is generally not the dominant source of \ion{C}{1} --- though it may make a significant contribution in some individual cases.}

\item{For a small number of component groups with independent estimates for $n_e$ and $n_{\rm H}$, the resulting fractional ionization is often much greater than the value 2 $\times$ 10$^{-4}$ that would be expected if most of the electrons come from photoionization of carbon and other heavy elements --- suggesting that $n_e$ has been overestimated (if only radiative recombination is assumed) and/or that hydrogen may be slightly ionized in predominantly neutral clouds.
Inclusion of grain-assisted recombination can reduce the apparent discrepancies in some lines of sight.}

\item{For that same small sample, the average electron density for carbon, sodium, and potassium (assuming only radiative recombination) appears to be roughly constant --- at $n_e$ $\sim$ 0.14$\pm$0.07 cm$^{-3}$ (mean $\pm$ standard deviation) --- independent of the local hydrogen density $n_{\rm H}$.
If grain-assisted recombination is included, the resulting $n_e$ appear to be smaller for larger $n_{\rm H}$ --- reflecting the increased effects of charge exchange at higher densities.
Observed relationships between the column densities of various neutral and singly ionized species and $N$(H) are difficult to understand if $n_e$ is either independent of or inversely proportional to $n_{\rm H}$.}

\item{Relationships between ratios of the column densities of \ion{Ca}{1}, \ion{K}{1}, \ion{Na}{1}, CN, and CH suggest that the depletion of calcium has a much weaker dependence on density than that inferred from previous comparisons based on more limited data.}

\end{itemize}

It has been common practice to assume photoionization equilibrium to estimate the electron density in diffuse interstellar clouds, then to assume a fractional ionization equal to the gas-phase carbon abundance to estimate $n_{\rm H}$.
The results discussed above indicate that such estimates must be viewed with caution --- especially if they are based on only one ratio $N$(\ion{X}{1})/$N$(\ion{X}{2}).
Because lines from both \ion{Ca}{1} and \ion{Ca}{2} are available in the optical, the Ca ionization balance is often used --- but the resulting $n_e$ can be higher than the values inferred from other elements by nearly an order of magnitude.
(And the ``true'' values from calcium would be even higher if \ion{Ca}{2} is more widely distributed than \ion{Ca}{1}.)
Even if other elements are used to determine $n_e$, the resulting fractional ionization (using $n_{\rm H}$ determined from \ion{C}{1} fine-structure excitation) can still be significantly larger than the gas-phase abundance of carbon if only radiative recombination is considered.
Inclusion of grain-assisted recombination appears to yield more consistent results for $n_e$ and $n_e$/$n_{\rm H}$ in some cases, but not for every element or for every line of sight.
It may prove interesting to revisit some of the possible correlations explored in this paper when more detailed and extensive abundances for the various atomic and molecular species become available for more {\it individual} interstellar clouds.
High-resolution observations of lines from many different neutral/first-ion pairs, for additional sightlines sampling different environmental conditions, may help us to understand the various processes affecting the ionization balance in diffuse clouds.

\acknowledgments

We acknowledge use of various routines obtained from the archive of statistical software maintained at Penn State (http://www.astro.psu.edu/statcodes).
We are grateful D. Willmarth (KPNO), D. Doss (McDonald), and R. P. Butler (AAO) for their help in setting up the spectrographs; to K. Pan and S. Federman for obtaining several of the coud\'{e} feed spectra; and to the anonymous referee for a careful reading of the original manuscript and for suggestions that have improved the paper.
Support for this work has been provided by NASA LTSA grant NAG5-3228 and by NASA grant GO-2251.01-87A from the Space Telescope Science Institute, which is operated by AURA, Inc., under NASA contract NAS5-26555.

\appendix

\section{High-Resolution Spectra of Fe I}
\label{sec-fe1}

During our 1997 March AAT run, we also obtained high-resolution spectra (FWHM $\sim$ 0.32 km~s$^{-1}$) with the UHRF of the weak \ion{Fe}{1} line at 3719\AA\ for nine lines of sight; Barlow et al. (1995) obtained similar spectra for $\zeta$ Oph.
The instrumental setup, data reductions, and analysis were very similar to those described above for the \ion{Ca}{1} spectra.
The normalized \ion{Fe}{1} spectra for six of the sightlines are shown in Figure~\ref{fig:fe1}, together with the profiles of either \ion{Ca}{1} or \ion{K}{1}. 
The measured \ion{Fe}{1} equivalent widths and the component structures derived from fits to the line profiles are given in Table~\ref{tab:fe1}; equivalent widths from the few previous studies to have measured this line are given in a footnote to the table.
We have used $f$ = 0.0413 (Morton 1991) in deriving the column densities.
Examination of the profiles in Figure~\ref{fig:fe1} indicates that the component structures seen in \ion{Fe}{1} are very similar to those for \ion{Ca}{1}; the \ion{Fe}{1} $b$-values are similar to those found for the corresponding components in \ion{Ca}{1} and \ion{K}{1}.
There are obvious variations in the ratio $N$(\ion{Fe}{1})/$N$(\ion{Ca}{1}), however (compare $\zeta$ Ori with $\delta$ Sco, $\sigma$ Sco, and $\rho$ Oph) --- reminiscent of the variations in relative $n_e$ derived from iron and calcium in different sightlines (\S~\ref{sec-relne} and Fig.~\ref{fig:relne}).

In Figure~\ref{fig:fe1ca1}, we plot $N$(\ion{Fe}{1}) vs. $N$(\ion{Ca}{1}) for all lines of sight with data for both species (see below).
The slope of the best-fit (solid) line is about 1.13$\pm$0.14 --- marginally steeper than that for a linear relationship between the two species, and perhaps suggestive of slightly less severe depletion of iron, relative to that of calcium, in the higher column density sightlines.
The scatter is about 0.21 dex, similar to that for $N$(\ion{Ca}{1}) vs. $N$(\ion{K}{1}).

\section{Column Densities for Ca I, Ca II, Fe I, and S I}
\label{sec-app}

In Table~\ref{tab:coldens}, we list total line-of-sight column densities for \ion{Ca}{1}, \ion{Ca}{2}, \ion{Fe}{1}, and/or \ion{S}{1} for 135 sightlines for which data for \ion{Ca}{1} are available.
An asterisk in the third column denotes sightlines included in the appendix to WH, where similar data for H, H$_2$, \ion{Na}{1}, \ion{K}{1}, \ion{Li}{1}, \ion{C}{1}, and CH are collected.
We have used $f$-values from Morton (1991), with updates for some lines as listed by Welty et al. (1999b), for determining the column densities.
%A database of column density measurements for all these species (plus several others) is being maintained by DEW.

{\bf \ion{Ca}{1}:}  The \ion{Ca}{1} column densities for sightlines observed for this paper are based on the fits to the high-resolution line profiles.
Values for some more heavily reddened sightlines are from a high-resolution survey of optical absorption lines toward stars in the {\it FUSE} PI team program studying H$_2$ in translucent cloud sightlines (Welty et al., in prep.; see Rachford et al. 2002).
Values for other sightlines were derived from equivalent widths of the $\lambda$4226 line (see references); thirty of those were measured from high S/N spectra obtained with the ARC echelle spectrograph (FWHM $\sim$ 8 km~s$^{-1}$) for a survey of diffuse interstellar bands (J. Thorburn, priv. comm.).
For W$_{\lambda}$ $\le$ 15 m\AA, we have assumed that the lines are optically thin, and have given lower limits for stronger lines.

{\bf \ion{Ca}{2}:}  We have only included \ion{Ca}{2} column densities derived from high-resolution spectra of the $\lambda$3933 K line (FWHM $\la$ 2 km~s$^{-1}$; mostly from WMH), as the \ion{Ca}{2} lines can be strong and complex where \ion{Ca}{1} is detected.

{\bf \ion{Fe}{1}:}  For \ion{Fe}{1}, very few high-resolution optical spectra are available (this paper; Barlow et al. 1995).
We include values obtained from lower resolution spectra (e.g., {\it HST}/GHRS echelle spectra of the lines at 2484 and 2523 \AA), as the measured absorption lines are generally relatively weak (e.g., Welty et al. 1999b). 
Forty-one of the values were derived from equivalent widths of the \ion{Fe}{1} lines at 3719 and 3859 \AA, obtained from the ARC diffuse band survey spectra noted above (J. Thorburn, priv. comm.).

{\bf \ion{S}{1}:}  For \ion{S}{1}, few results from high-resolution UV spectra have been reported, and there are no lines at optical wavelengths.
We have used component structures derived from higher resolution \ion{K}{1} and/or \ion{Na}{1} spectra (when available) to model the measured \ion{S}{1} equivalent widths (typically for the lines at 1295, 1425, and 1807 \AA; e.g., Welty et al. 1999b); otherwise, we have scaled column densities in the original references to the current $f$-values.

While we have not listed uncertainties for the column densities in Table~\ref{tab:coldens} (or in Table~7 of WH), where possible we have tabulated, derived, or estimated uncertainties for those column densities for use in the correlation and regression analyses\footnotemark.
For column densities derived from fairly weak absorption lines (W$_{\lambda}$ $\la$ 15 m\AA), the column density uncertainties generally correspond to the uncertainties in the equivalent widths --- which are often of order 10--20\% (0.04--0.08 dex) even for confidently detected lines.
Most values for \ion{Ca}{1}, \ion{Fe}{1}, and \ion{Li}{1} --- and some values for \ion{Na}{1}, \ion{K}{1}, \ion{Ca}{2}, and CH --- are based on such weak lines.
For column densities derived from very strong lines (many of the values for \ion{Na}{1}, \ion{K}{1}, and \ion{Ca}{2}), the formal uncertainties in the total column densities derived from fits to high-resolution, high-S/N spectra are often $\la$ 0.02 dex.
The true uncertainties are likely to be somewhat larger, however, as the fits {\it assume} the number of components present and (in many cases) fix some of the $b$-values to facilitate convergence.
Moreover, component structure information from other, similarly distributed species also may be used to fit very strong lines and/or absorption-line profiles observed at slightly lower resolution.
While the effects of those assumptions are difficult to quantify, in some cases the column densities can be compared with those derived from weak lines of the same species.
For example, \ion{K}{1} column densities derived from fits to the strong $\lambda$7698 line were within 0.05 dex of those obtained from the weak $\lambda$4044 line for three of the five sightlines with reliable data for both lines (WH).
We therefore have adopted a minimum uncertainty of 0.03--0.04 dex for column densities derived from fits to relatively strong lines (for all species).
For the species considered in this paper, the typical and adopted minimum uncertainties are:  \ion{C}{1} (0.2/0.1 dex), \ion{Na}{1} (0.04/0.04 dex), \ion{S}{1} (0.1/0.05 dex), \ion{K}{1} (0.03/0.03 dex), \ion{Ca}{1} (0.06/0.03 dex), \ion{Ca}{2} (0.03/0.03 dex), \ion{Fe}{1} (0.07/0.04 dex), \ion{H}{1} (0.1/0.05 dex), H$_2$ (0.2/0.05 dex), CH (0.04/0.03 dex), CH$^+$ (0.05/0.03 dex), and CN (0.09/0.04 dex).
\footnotetext{In some cases, the original references (including our previous surveys) do not give uncertainties for either the measured equivalent widths or the derived column densities.
In other cases, it is not clear what effects were included (e.g., photon noise, continuum placement, background uncertainties), or exactly how the listed uncertainties were calculated --- making it difficult to compare values quoted in different studies.
An updated compilation of column densities (with estimated uncertainties) may be found at http://astro.uchicago.edu/home/web/welty/coldens.html.
Compilations of equivalent widths measured from high-resolution optical spectra (with uncertainties) may be found at http://astro.uchicago.edu/home/web/welty/ew-atom.html (for \ion{Na}{1}, \ion{K}{1}, \ion{Ca}{1}, and \ion{Ca}{2}) and http://astro.uchicago.edu/home/web/welty/ew-mol.html (for CN $\lambda$3875, CH $\lambda$4300, and CH$^+$ $\lambda$4232).}

\pagebreak

\begin{figure}
\figurenum{1a}
\epsscale{0.9}
\plotone{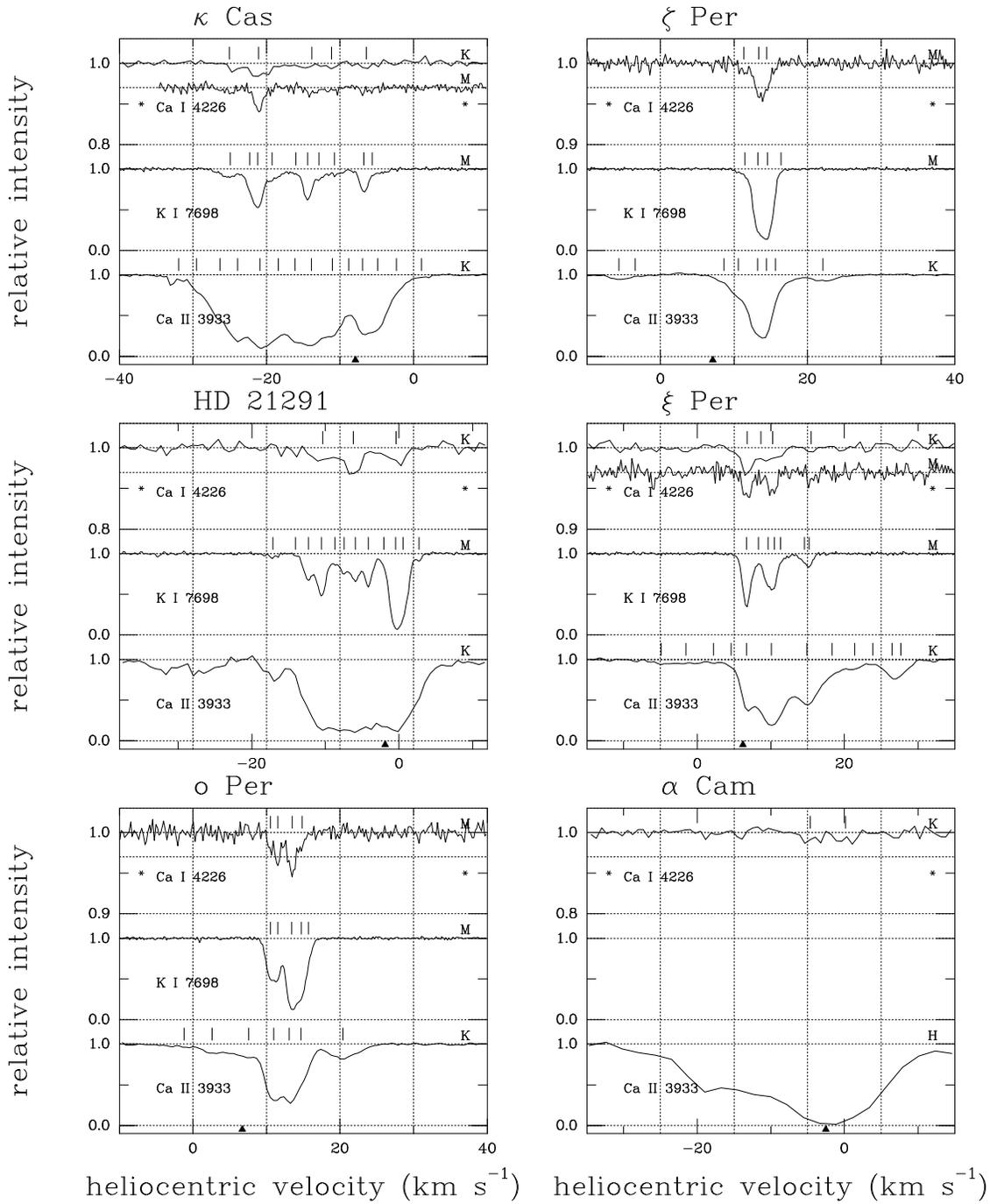}
\caption{The interstellar Ca~I, K~I (or Na~I), and Ca~II profiles.
The vertical scale has been expanded in some cases to show weaker lines (especially of Ca~I) more clearly (noted by an asterisk near both vertical axes).
The sources of the spectra have been noted at the right, just above the continuum [M = McDonald 2.7m coud\'{e} (FWHM = 0.53--0.56 km~s$^{-1}$); K = KPNO coud\'{e} feed (FWHM = 1.2--1.5 km~s$^{-1}$); A = AAT UHRF (FWHM = 0.3--0.4 km~s$^{-1}$); E = ESO 3.6m + CES (FWHM = 1.2 km~s$^{-1}$); H = Hobbs (PEPSIOS; FWHM = 1--2 km~s$^{-1}$)].
Tick marks above the spectra indicate the components found in fitting the profiles.
The solid triangles denote $v_{\rm LSR}$ = 0 km~s$^{-1}$.}
\label{fig:spec1}
\end{figure}

\begin{figure}
\figurenum{1b}
\epsscale{0.9}
\plotone{f1b.eps}
\caption{}
\label{fig:spec2}
\end{figure}
 
\begin{figure}
\figurenum{1c}
\epsscale{0.9}
\plotone{f1c.eps}
\caption{}        
\label{fig:spec3}
\end{figure}
 
\begin{figure}
\figurenum{1d}
\epsscale{0.9}
\plotone{f1d.eps}
\caption{}   
\label{fig:spec4}
\end{figure}
 
\begin{figure}
\figurenum{1e}
\epsscale{0.9}
\plotone{f1e.eps}
\caption{}
\label{fig:spec5}
\end{figure}
 
\begin{figure}
\figurenum{2}
\epsscale{0.9}
\plotone{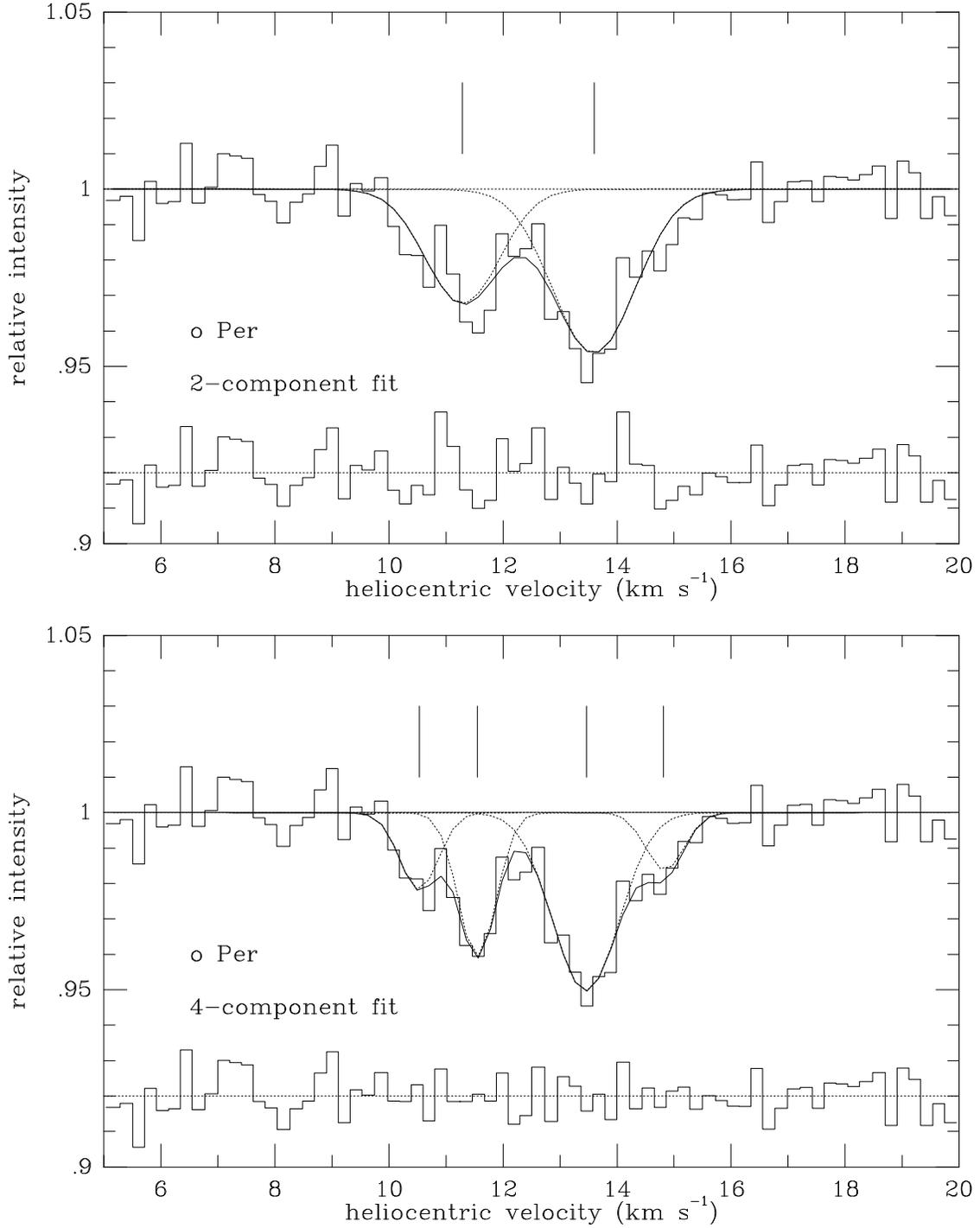}
\caption{Fits to the Ca~I profile observed toward o~Per.
The top panel shows a two-component fit --- which is not adequate; the bottom panel shows the adopted four-component fit.
In both panels, the individual components (with velocities given by the tick marks) are shown by dotted lines; the composite fits are shown by the smooth solid line; the data are shown by the upper solid histogram; the residuals (data minus fit) are shown by the lower solid histogram.}
\label{fig:fit}
\end{figure}

\begin{figure}
\figurenum{3}
\epsscale{0.7}
%\plotone{ca1vsk1_all.eps}
\plotone{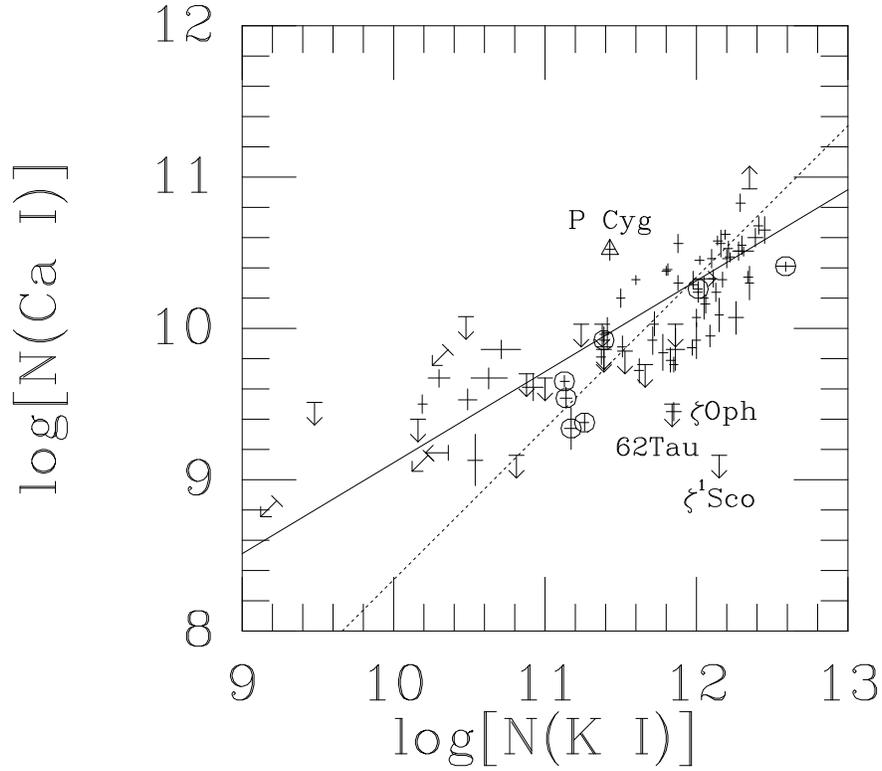}
\caption{N(Ca~I) vs. N(K~I).
Open circles denote Sco-Oph sightlines; open triangles denote other ``discrepant'' sightlines identified by WH.
The size of each plus sign indicates the $\pm$1$\sigma$ uncertainties in each quantity.
The slope of the best-fit line (solid) is 0.60$\pm$0.06 --- significantly shallower than that for a linear relationship (dotted line) --- which may reflect increasing severity of Ca depletion in higher column density (higher mean density?) sightlines.}
\label{fig:ca1k1}
\end{figure}

\begin{figure}
\figurenum{4}
\epsscale{0.9}
%\plotone{cakvsca.eps}
\plotone{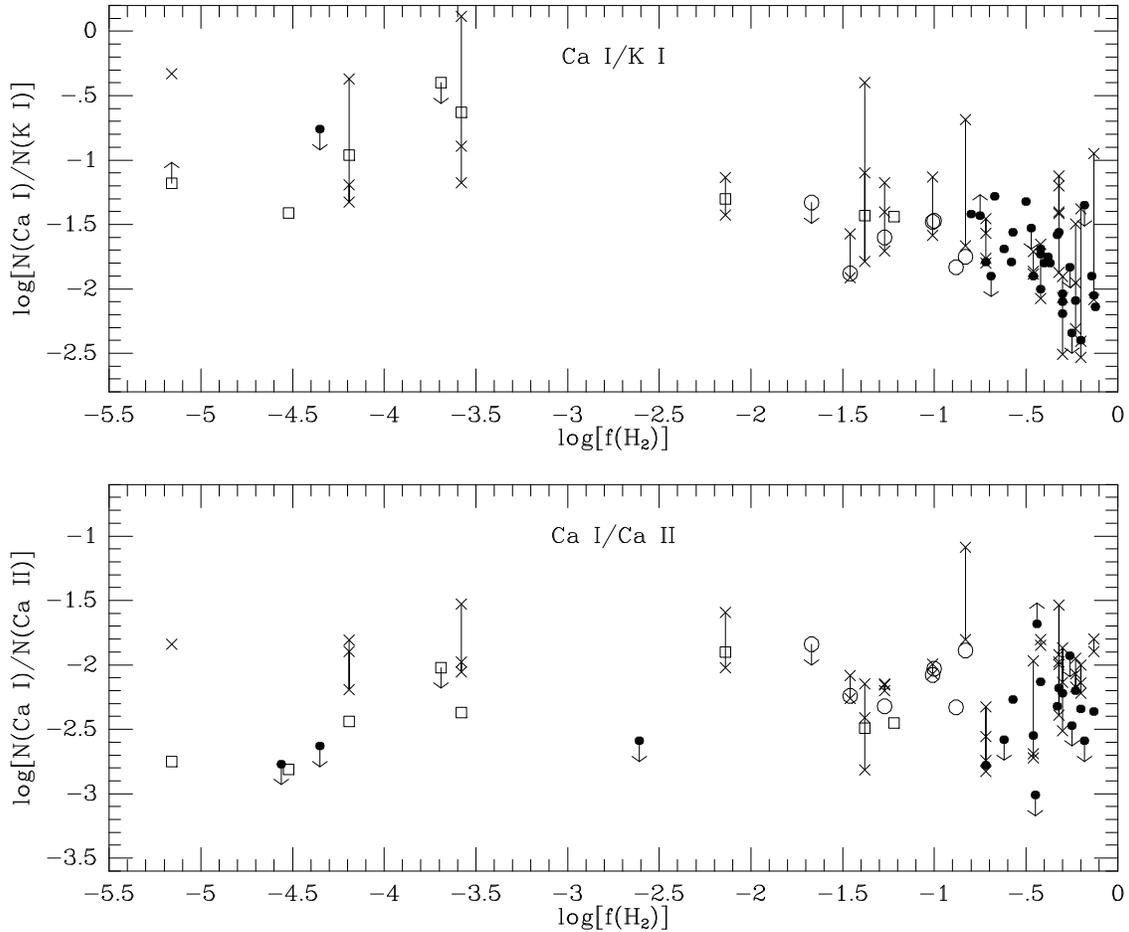}
\caption{N(Ca~I)/N(K~I) and N(Ca~I)/N(Ca~II) vs. $f$(H$_2$).
Open circles denote Sco-Oph sightlines; open squares denote Orion sightlines; filled circles denote other sightlines.
Crosses (connected by lines) denote values for (some) individual components, assumed to have the same $f$(H$_2$) as the average for the line of sight.
The total line-of-sight ratios probably underestimate the true values where Ca~I is detected over a much smaller velocity range than K~I and/or Ca~II.
N(Ca~I)/N(K~I) is smaller for larger $f$(H$_2$) --- consistent with the ratio being an indicator for D(Ca).
N(Ca~I)/N(Ca~II) shows no obvious dependence on $f$(H$_2$) --- so either $n_e$/($\Gamma$/$\alpha$) is independent of $n_{\rm H}$ or $f$(H$_2$) is not a reliable indicator of $n_{\rm H}$ or photoionization equilibrium does not hold.
Typical 1$\sigma$ uncertainties in both ratios are of order 0.07 dex.}
\label{fig:cakvsfh2}
\end{figure}

\begin{figure}
\figurenum{5}
\epsscale{0.8}
%\plotone{xvsca.eps}
\plotone{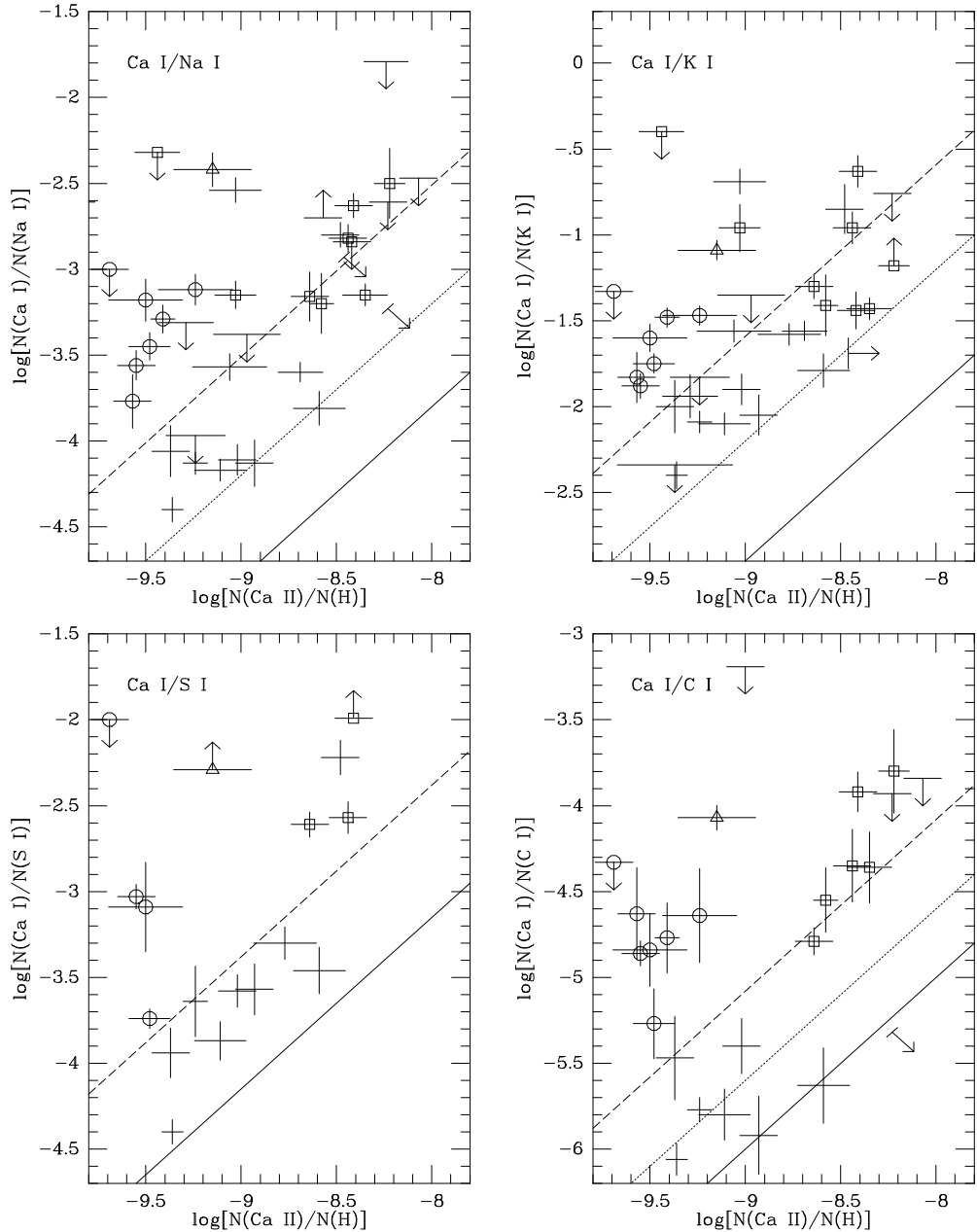}
\caption{N(Ca~I)/N(X~I) vs. N(Ca~II)/N(H) for various elements X.
Open circles denote Sco-Oph sightlines; open squares denote Orion sightlines; open triangles denote main LMC components toward SN 1987A.
The size of each plus sign indicates the $\pm$1$\sigma$ uncertainties in each quantity.
The solid lines give the expected (linear) relationships if X is undepleted (assuming the photoionization and radiative recombination rates in Table 3); the dotted lines give the relationships for typical depletions ($-$0.4 dex for C, $-$0.6 for Na, $-$0.7 for K, 0.0 for S); for comparison, the dashed lines show the fits to the data with the slopes fixed at unity.}
\label{fig:xvsca}
\end{figure}

\begin{figure}
\figurenum{6}
\epsscale{1.0}
\plottwo{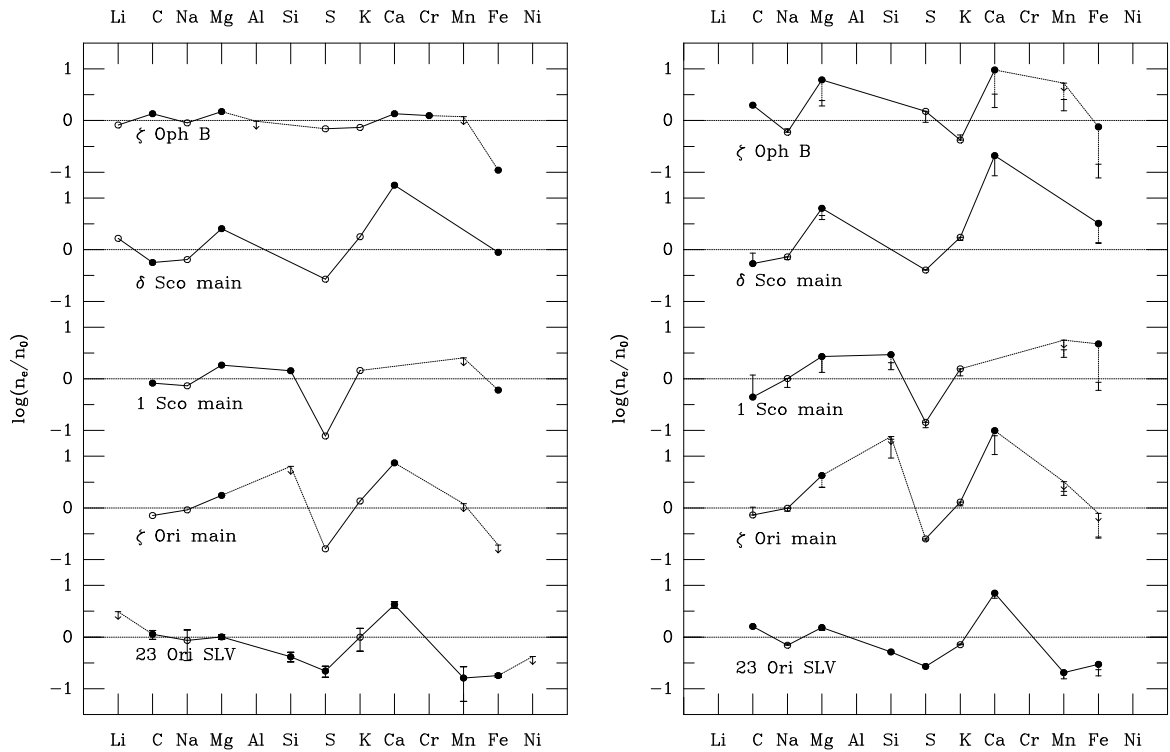}{f6b.eps}
\caption{({\it left}) Logarithm of $n_e$ = N(X~I)/N(X~II) * ($\Gamma$/$\alpha_r$), for $T$ = 100 K and the WJ1 radiation field, relative to $n_0$ (the average value for C, Na, and K), for the main components toward five stars (Table 6).
Open circles denote cases where X~II was estimated based on N(H) and typical depletions; representative 1~$\sigma$ error bars are shown for 23~Ori.
Note the similarities between 23~Ori and $\zeta$~Ori, and between 1~Sco and $\delta$~Sco --- and the differences between those two pairs and $\zeta$~Oph.
The points for S, Mn, Fe, and Ni are often low; Ca is often high.
({\it right}) The same plot, except that charge exchange with small grains is included (Weingartner \& Draine 2001).
The circles denote the $n_e$/$n_0$ values for the adopted set of sticking parameters $s$ (see text); the attached ``error bar'' denotes the range between all $s$=0 and all $s$=1.
The assumed depletions of Na and K are less severe by $\sim$ 0.2 dex.
The element-to-element differences are generally similar to the case where radiative recombination is dominant, however.}
\label{fig:relne}
\end{figure}

\begin{figure}
\figurenum{7}
\epsscale{1.0}
\plotone{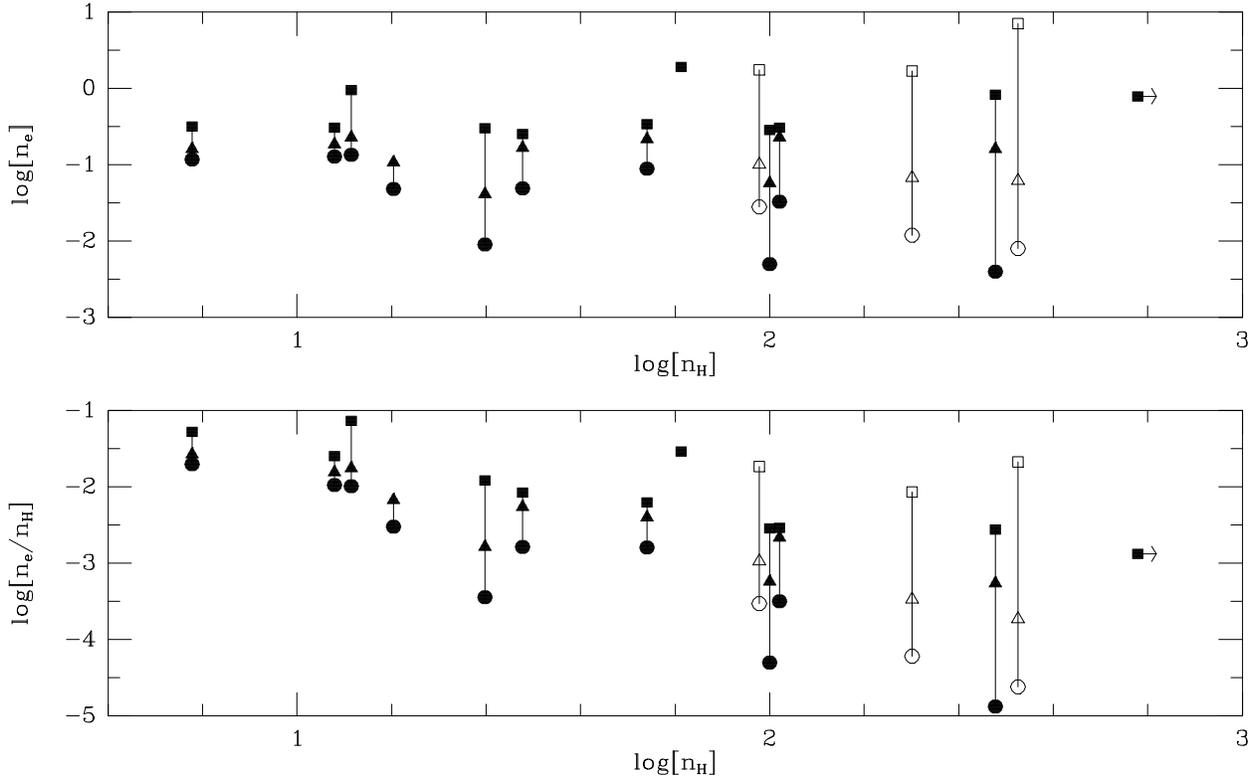}
\caption{Electron density $n_e$ (top) and fractional ionization $n_e$/$n_{\rm H}$ (bottom) versus local hydrogen density $n_{\rm H}$.
Squares are values derived from calcium; circles and triangles are average values derived from carbon, sodium, and potassium --- with and without grain-assisted recombination, respectively (see Table 7).
Open symbols denote the three Scorpius sightlines, where the radiation field was assumed stronger by a factor of 5.
Typical uncertainties are less than about 0.3 dex for log($n_{\rm H}$), but may be somewhat larger for log($n_e$).
Note that $n_e$(rad) (CNaK) (triangles) appears to be independent of $n_{\rm H}$, within those uncertainties.}
\label{fig:nevsnh}
\end{figure}

\begin{figure}
\figurenum{8}
\epsscale{0.5}
%\plotone{dissrec.eps}
\plotone{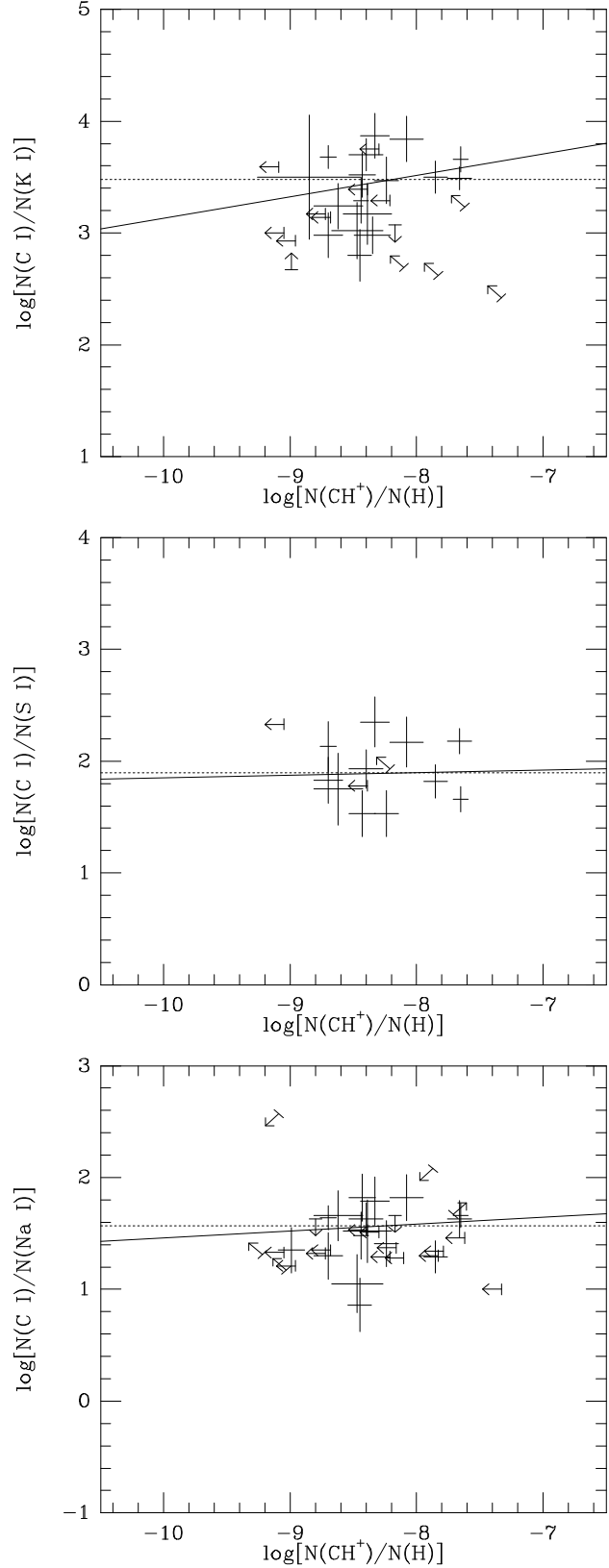}
\caption{$N$(C~I)/$N$(X~I) vs. $N$(CH$^+$)/$N$(H), for X = Na, K, and S.
The size of each plus sign indicates the $\pm$1$\sigma$ uncertainties in each quantity.
Dotted lines show the mean values of the $N$(C~I)/$N$(X~I) ratios; solid lines show the correlation fits, with slopes 0.0--0.2 (Table 5).
Linear correlations (with slope = 1.0) would have been expected if dissociative recombination of CH$^+$ dominates the production of C~I.}
\label{fig:dissrec}
\end{figure}

\begin{figure}
\figurenum{9}
\epsscale{0.5}
%\plotone{cfs1.eps}
\plotone{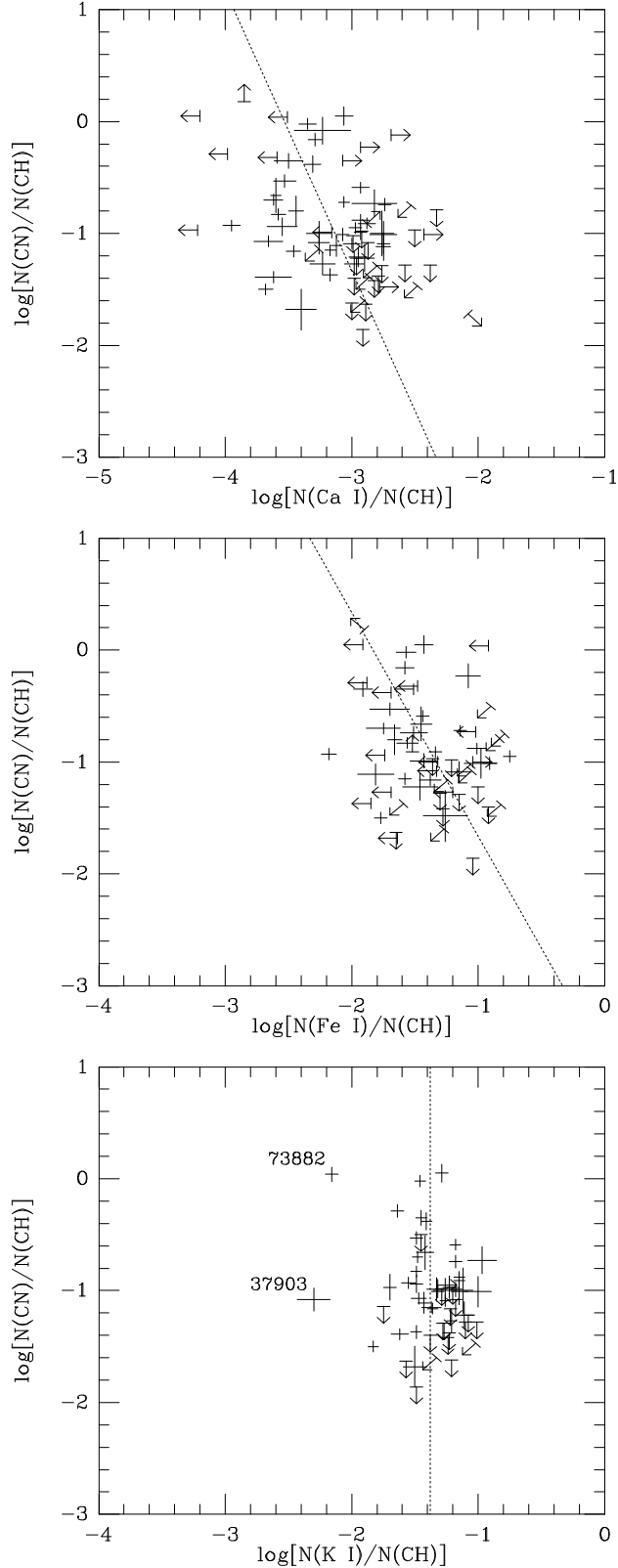}
\caption{$N$(CN)/$N$(CH) vs. $N$(X~I)/$N$(CH), for X = Ca, K, and Fe (following Cardelli et al. 1991).
The size of each plus sign indicates the $\pm$1$\sigma$ uncertainties in each quantity.
For Ca and Fe, there may be inverse relationships between the two ratios, but with steeper slopes ($\la$ $-$2.5 and $\la$ $-$2.0; dotted lines) than the $-$1 found by Cardelli et al. for Ca --- suggesting that the depletions of Ca and Fe do not depend as strongly on the local density as $n_{\rm H}^{-3}$.
For K, the slope is essentially infinite, as $N$(K~I)/$N$(CH) is roughly constant for most Galactic sightlines (WH).}
\label{fig:cfs1}
\end{figure}

\begin{figure}
\figurenum{10}
\epsscale{0.5}
%\plotone{cfs2.eps}
\plotone{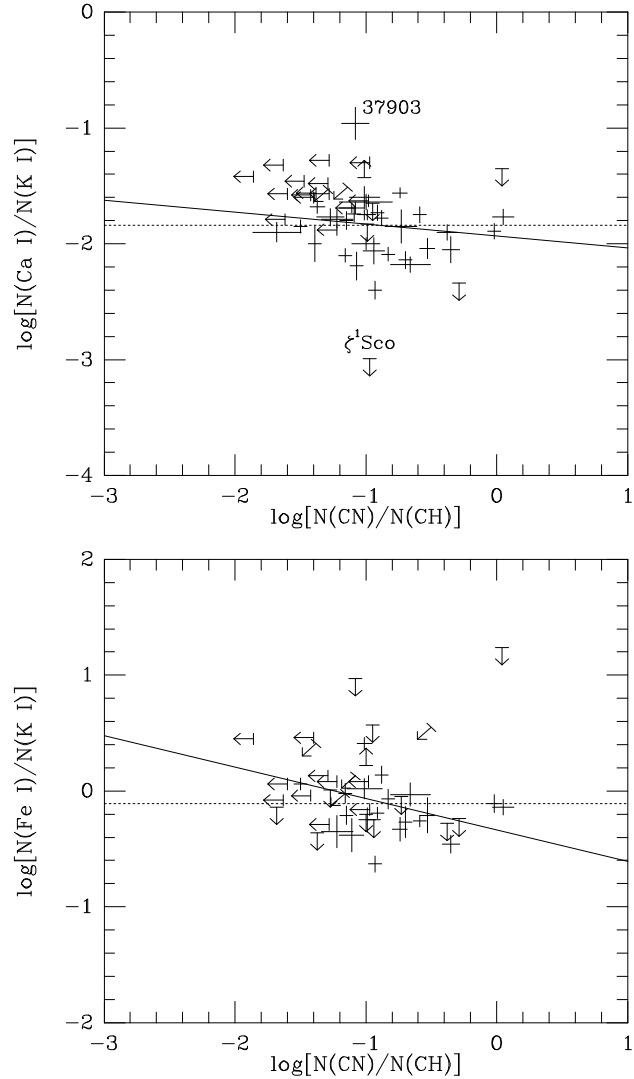}
\caption{$N$(Ca~I)/$N$(K~I) and $N$(Fe~I)/$N$(K~I) vs. $N$(CN)/$N$(CH).
The size of each plus sign indicates the $\pm$1$\sigma$ uncertainties in each quantity.
Weak trends, with slightly negative slopes, may be present (especially if the limits are considered).
If $n$(CN)/$n$(CH) $\sim$ $N$(CN)/$N$(CH) is proportional to $n_{\rm H}^{2}$, and if the depletion of K does not depend on density, then those weak trends would suggest that the depletions of Ca and Fe go (at most) roughly as $n_{\rm H}^{-0.5}$.}
\label{fig:cfs2}
\end{figure}

\begin{figure}
\figurenum{11}
\epsscale{0.9}
\plotone{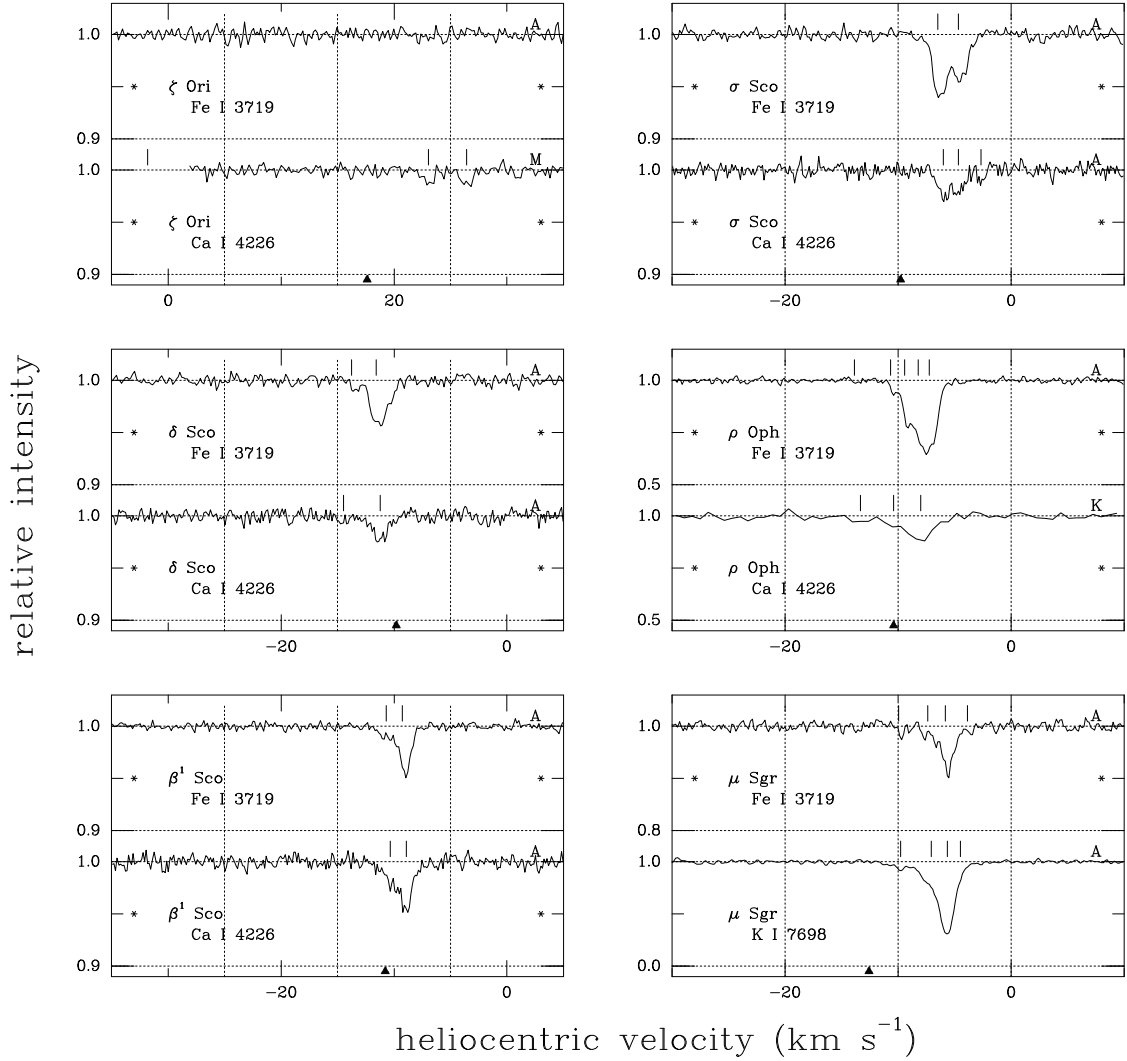}
\caption{The interstellar Fe~I $\lambda$3719 profiles obtained with the AAT/UHRF (FWHM $\sim$ 0.3 km~s$^{-1}$) toward six stars.
The corresponding Ca~I or K~I profiles are given for comparison.
Tick marks indicate components found in fitting the profiles.
The vertical scales are the same for Fe~I and Ca~I in all cases; note the variations in relative strength of the two lines.} 
\label{fig:fe1}
\end{figure}

\begin{figure}
\figurenum{12}
\epsscale{0.7}
%\plotone{fe1vsca1_all.eps}
\plotone{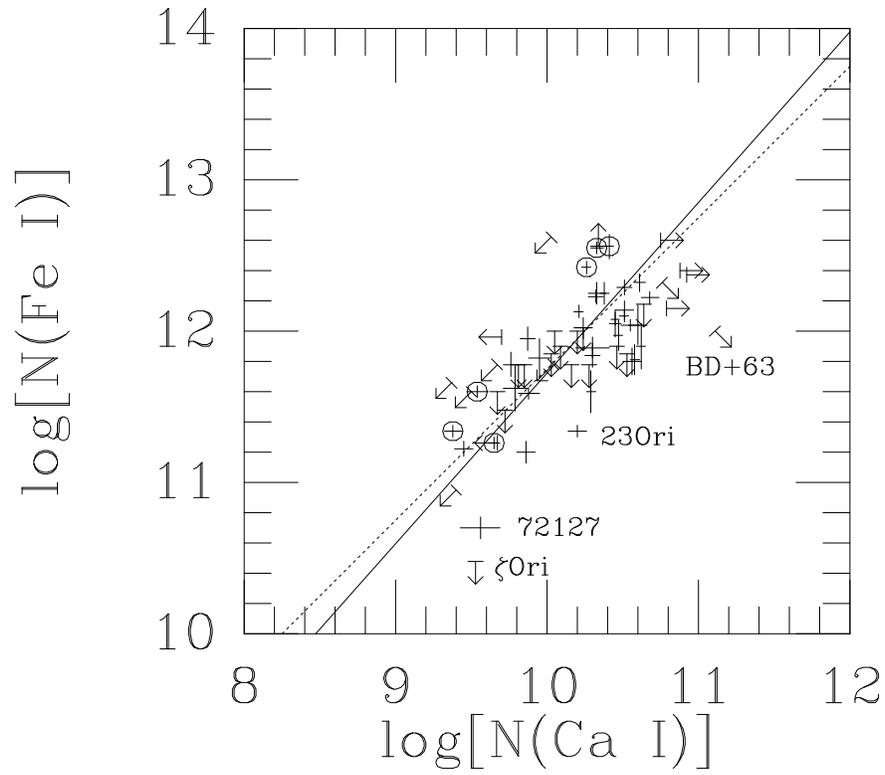}
\caption{N(Fe~I) vs. N(Ca~I).
The size of each plus sign indicates the $\pm$1$\sigma$ uncertainties in each quantity.
Open circles denote Sco-Oph sightlines.
The slope of the best-fit line (solid) is 1.13$\pm$0.14 --- consistent with a linear (or perhaps slightly steeper) relationship between the two species (dotted line).}
\label{fig:fe1ca1}
\end{figure}

\newpage

\begin{deluxetable}{lrcrcrcl}
\tabletypesize{\scriptsize}
\tablecolumns{8}
\tablenum{1}
\tablecaption{Ca I Equivalent Widths\tablenotemark{a} \label{tab:stel}}
\tablewidth{0pt}
 
\tablehead{
\multicolumn{1}{l}{Star}&
\multicolumn{2}{c}{-~-~-~CF~-~-~-}&
\multicolumn{2}{c}{-~-~-~McD~-~-~-}&
\multicolumn{2}{c}{-~-~-~UHRF~-~-~-}&
\multicolumn{1}{l}{Previous\tablenotemark{b}}\\
\multicolumn{1}{c}{ }&
\multicolumn{1}{c}{S/N}&
\multicolumn{1}{c}{W$_\lambda$}&
\multicolumn{1}{c}{S/N}&
\multicolumn{1}{c}{W$_\lambda$}&
\multicolumn{1}{c}{S/N}&
\multicolumn{1}{c}{W$_\lambda$}&
\multicolumn{1}{c}{ }\\
\multicolumn{1}{c}{ }&
\multicolumn{1}{c}{ }&
\multicolumn{1}{c}{(m\AA)}&
\multicolumn{1}{c}{ }&
\multicolumn{1}{c}{(m\AA)}&
\multicolumn{1}{c}{ }&
\multicolumn{1}{c}{(m\AA)}&
\multicolumn{1}{c}{ }}
\startdata
$\kappa$ Cas   & 205 & 3.2$\pm$1.3 & 155 & 2.5$\pm$1.7 & \nodata & \nodata &
     $<$0.9 (1) \\
HD 21291       & 135 & 5.7$\pm$1.4 & \nodata & \nodata & \nodata & \nodata &
     10.3$\pm$1.5 (2) \\
o Per          & \nodata & \nodata & 160 & 2.0$\pm$0.4 & \nodata & \nodata &
     1.5$\pm$0.9 (3); $<$2.0 (4) \\
$\zeta$ Per    & \nodata & \nodata & 190 & 1.6$\pm$0.4 & \nodata & \nodata &
     $<$0.9 (1); 0.9$\pm$0.6 (3); $<$2.4 (4) \\
$\xi$ Per      & 220 & 1.6$\pm$0.5 & 165 & 1.5$\pm$0.8 & \nodata & \nodata &
     $<$1.1 (1); 2.3$\pm$0.6 (2) \\
$\alpha$ Cam   & 120 & 1.9$\pm$1.0 & \nodata & \nodata & \nodata & \nodata &
     2.4$\pm$0.4 (5) \\
23 Ori         &  65 & 4.1$\pm$1.3 & \nodata & \nodata & \nodata & \nodata &
     4.7$\pm$0.9 (2); 5.6$\pm$0.8 (5); 3.2$\pm$1.6 (6) \\
$\lambda$ Ori  & 120 & 2.1$\pm$0.8 & 190 & 2.9$\pm$0.6 & \nodata & \nodata &
     3.8$\pm$0.6 (5) \\
\enddata
\tablenotetext{a}{Uncertainties and limits are 2$\sigma$.}
\tablenotetext{b}{Equivalent width (reference) for previous data.}
\tablenotetext{c}{Not including component at $-$2 km~s$^{-1}$.}
\tablerefs{1 = Hobbs 1971; 2 = Welty (unpubl. CF camera 5); 3 = White 1973; 4 = Allen 1994; 5 = Federman \& Hobbs 1983; 6 = Lambert \& Danks 1986; 7 = Snow et al. 1986; 8 = Shulman et al. 1974; 9 = Gredel et al. 1993; 10 = Chaffee \& Dunham 1979}
\end{deluxetable}

\newpage

\begin{deluxetable}{lcrrrc}
\tabletypesize{\scriptsize}
\tablecolumns{6}
\tablenum{2}
\tablecaption{Ca I Component Structures\label{tab:comp}}
\tablewidth{0pt}
 
\tablehead{
\multicolumn{1}{l}{Star}&
\multicolumn{1}{c}{Obs\tablenotemark{a}}&
\multicolumn{1}{c}{Comp}&
\multicolumn{1}{c}{$v_\odot$}&
\multicolumn{1}{c}{$N$}&
\multicolumn{1}{c}{$b$\tablenotemark{b}}\\
\multicolumn{1}{c}{ }&
\multicolumn{1}{c}{ }&
\multicolumn{1}{c}{ }&
\multicolumn{1}{c}{(km~s$^{-1}$)}&
\multicolumn{1}{c}{(10$^{9}$ cm$^{-2}$)}&
\multicolumn{1}{c}{(km~s$^{-1}$)}}
\startdata
$\kappa$ Cas   &M& 1&$-$25.07&    1.3& 1.27 \\              
               & & 2&$-$21.06&    4.6& 0.80 \\              
               & & 3&$-$13.86&    1.4& 0.91 \\              
               & & 4&$-$11.19&    0.8&(1.00)\\              
               & & 5& $-$6.44&    0.9& 1.02 \\              
               & &  &        &       &      \\
HD 21291       &K& 1&$-$10.36&    6.1& 1.86 \\
               & & 2& $-$6.20&    9.3& 1.19 \\
\enddata
\tablenotetext{a}{Observatory:  M = McDonald (FWHM = 0.53 km~s$^{-1}$); A = AAO (FWHM = 0.3 km~s$^{-1}$); K = KPNO (FWHM = 1.3--1.5 km~s$^{-1}$).}
\tablenotetext{b}{Values of $b$ in square brackets were fixed to facilitate convergence of the fits, but are well-determined.  Values of $b$ in parenthese were also fixed, but are less well-determined.}
\end{deluxetable}

\begin{deluxetable}{lcrcccccc}
\tabletypesize{\scriptsize}
\tablecolumns{9}
\tablenum{3}
\tablecaption{Abundances and Atomic Data\label{tab:atom}}
\tablewidth{0pt}
 
\tablehead{
\multicolumn{1}{l}{Element}&
\multicolumn{1}{c}{A$_{\sun}$\tablenotemark{a}}&
\multicolumn{1}{c}{$\chi_{\rm ion}$}&
\multicolumn{1}{c}{$\Gamma$ $\times$ 10$^{11}$\tablenotemark{b}}&
\multicolumn{1}{c}{$\Delta$(log $\Gamma$)\tablenotemark{c}}&
\multicolumn{1}{c}{$\alpha_r$ $\times$ 10$^{12}$\tablenotemark{b}}&
\multicolumn{1}{c}{$\Gamma$/$\alpha_r$}&
\multicolumn{1}{c}{$\alpha_g$ $\times$ 10$^{14}$\tablenotemark{d}}&
\multicolumn{1}{c}{$\Gamma$/$\alpha_g$ $\times$ 10$^3$}\\
\multicolumn{1}{c}{ }&
\multicolumn{1}{c}{ }&
\multicolumn{1}{c}{(eV)}&
\multicolumn{1}{c}{(s$^{-1}$)}&
\multicolumn{1}{c}{ }&
\multicolumn{1}{c}{(cm$^3$s$^{-1}$)}&
\multicolumn{1}{c}{(cm$^{-3}$)}&
\multicolumn{1}{c}{(cm$^3$s$^{-1}$)}&
\multicolumn{1}{c}{(cm$^{-3}$)}}
\startdata
Li                 &3.31& 5.39& 34.  &\nodata& 9.6 & 35.4 &\nodata&\nodata\\
C                  &8.55&11.26& 20.  &$-$0.70& 8.3 & 24.  & 1.62 & 12.3  \\
Na                 &6.31& 5.14&  1.3 &$-$0.46& 5.8 &  2.2 & 0.93 &  1.40 \\
Mg                 &7.58& 7.65&  7.9 &$-$0.35& 5.9 & 13.5 & 1.18 &  6.70 \\
Al                 &6.48& 5.99&390.  &$-$0.37&15.4 &253.  &\nodata&\nodata\\ 
Si                 &7.55& 8.15&300.  &$-$0.39& 8.4 &357.  & 1.16 &259.   \\
S                  &7.27&10.36& 78.  &$-$0.64&10.5 & 74.  & 1.03 & 75.7  \\
K                  &5.13& 4.34&  5.6 &$-$0.49& 5.5 & 10.1 & 0.76 &  7.37 \\
\enddata
\tablenotetext{a}{Solar system abundances (logarithmic) from Anders \& Grevesse 1989; Grevesse \& Noels 1993 (H = 12.0).}
\tablenotetext{b}{P\'{e}quignot \& Aldrovandi 1986; photoionization rate $\Gamma$ for WJ1 radiation field; radiative recombination coefficient $\alpha_r$ for $T$ = 100 K.}
\tablenotetext{c}{Difference in log $\Gamma$ (edge minus center) for cloud of total A$_{\rm V}$ $\sim$ 1.0, for grain model 2 of Roberge et al. 1981.}
\tablenotetext{d}{Grain recombination coefficient (per H atom) from Weingartner \& Draine 2001; for $T$ = 100 K, $\psi$ = 400 K$^{1/2}$ cm$^3$.}
\tablenotetext{e}{Second entry is for Ca$^+$.}
\tablenotetext{f}{$\Gamma$ from Meyer \& Roth 1990; $\alpha_r$ estimated.}
\end{deluxetable}

\begin{deluxetable}{llrrrrrrc}
\tabletypesize{\scriptsize}
\tablecolumns{9}
\tablenum{4}
\tablecaption{Correlation Probabilities\label{tab:corr}}
\tablewidth{0pt}
 
\tablehead{
\multicolumn{1}{c}{Y}&
\multicolumn{1}{c}{X}&
\multicolumn{1}{c}{N$_{\rm tot}$\tablenotemark{a}}&
\multicolumn{1}{c}{N$_{\rm lim}$\tablenotemark{a}}&
\multicolumn{1}{c}{r\tablenotemark{b}}&
\multicolumn{1}{c}{P$_{\rm r}$\tablenotemark{b}}&
\multicolumn{1}{c}{P$_{\rm C}$\tablenotemark{c}}&
\multicolumn{1}{c}{P$_{\rm K}$\tablenotemark{d}}&
\multicolumn{1}{c}{P$_{\rm S}$\tablenotemark{e}}}
\startdata
Ca I      & K I       & 82 & 19 &    0.749 & 0.000 &\nodata& 0.000 & 0.000 \\
Fe I      & K I       & 51 & 18 &    0.768 & 0.000 & 0.000 & 0.000 & 0.000 \\
S I       & K I       & 19 &  3 &    0.926 & 0.000 & 0.000 & 0.000 & 0.000 \\
Fe I      & Ca I      & 59 & 23 &    0.664 & 0.000 &\nodata& 0.000 & 0.000 \\
 & \\
Ca I/Na I & Ca II/H   & 37 & 13 &    0.569 & 0.009 & 0.118 & 0.036 & 0.084 \\
Ca I/K I  & Ca II/H   & 33 &  7 &    0.539 & 0.012 & 0.020 & 0.007 & 0.010 \\
Ca I/S I  & Ca II/H   & 15 &  1 &    0.697 & 0.006 & 0.007 & 0.043 &(0.086)\\
\enddata
\tablenotetext{a}{N$_{\rm tot}$ = total number of points (detections plus limits); N$_{\rm lim}$ = number of points with limits for one or both values.}
\tablenotetext{b}{Linear correlation coefficient and associated probability of no correlation, using only detected points (Bevington 1969); average values considering either X or Y as the dependent variable.}
\tablenotetext{c}{Probability of no correlation from Cox's proportional hazard model, including limits (ASURV; Isobe et al. 1986).}
\tablenotetext{d}{Probability of no correlation from generalized Kendall's tau method, including limits (ASURV; Isobe et al. 1986).}
\tablenotetext{e}{Probability of no correlation from generalized Spearman's rank order method, including limits (ASURV; Isobe et al. 1986).
Values in parentheses are uncertain due to small number of points ($<$30).}
\end{deluxetable}

\begin{deluxetable}{llrccccc}
\tabletypesize{\scriptsize}
\tablecolumns{8}
\tablenum{5}
\tablecaption{Correlation Fits \label{tab:fits}}
\tablewidth{0pt}
 
\tablehead{
\multicolumn{8}{c}{log[$N$(Y)] = A + B $\times$ log[$N$(X)]}\\
\multicolumn{1}{c}{Y}&
\multicolumn{1}{c}{X}&
\multicolumn{1}{c}{A\tablenotemark{a}}&
\multicolumn{1}{c}{B\tablenotemark{b}}&
\multicolumn{1}{c}{RMS\tablenotemark{c}}&
\multicolumn{1}{c}{N\tablenotemark{d}}&
\multicolumn{1}{c}{Sample\tablenotemark{e}}&
\multicolumn{1}{c}{Program\tablenotemark{f}}}
\startdata
Ca I      & K I      &  $3.11\pm0.68$ & $0.60\pm0.06$ & 0.21 & 62 & 3/3 & R \\
          &          & $-1.66\pm0.32$ & (1.00)        & 0.23 & 61 & 3/3 & R \\
          &          &  $3.71\pm0.72$ & $0.55\pm0.06$ & 0.20 & 54 & 3/1 & R \\
          &          & $-1.66\pm0.31$ & (1.00)        & 0.22 & 53 & 3/1 & R \\
%& \\
Fe I      & K I      &  $2.31\pm1.09$ & $0.80\pm0.09$ & 0.19 & 32 & 3/3 & R \\
          &          &  $0.00\pm0.27$ & (1.00)        & 0.19 & 32 & 3/3 & R \\
          &          &  $1.14\pm1.58$ & $0.90\pm0.13$ & 0.17 & 26 & 3/1 & R \\
          &          & $-0.10\pm0.25$ & (1.00)        & 0.17 & 26 & 3/1 & R \\
\enddata
\tablenotetext{a}{Uncertainties in A were determined in the fits when B was varied, but are the standard deviations of the set of values log\{$N$(Y)/[$N$(X)]$^{\rm B}$\} when B was fixed.}
\tablenotetext{b}{Values in parentheses were fixed.}
\tablenotetext{c}{RMS distance of points from best-fit line.}
\tablenotetext{d}{Number of stars in sample.}
\tablenotetext{e}{n/m: n = 1 includes high resolution (FWHM $<$ 2 km~s$^{-1}$) data only; n = 3 includes also lower resolution data; m = 1 excludes Sco-Oph and other ``discrepant'' stars; m = 3 includes all stars.}
\tablenotetext{f}{R = regrwt (no limits; weighted by 1/$\sigma^2$); A = ASURV EM algorithm (limits; no weights).}
\end{deluxetable}

\begin{deluxetable}{lccccc}
\tabletypesize{\scriptsize}
\tablecolumns{6}
\tablenum{6}
\tablecaption{Relative Electron Densities\tablenotemark{a}\label{tab:relne}}
\tablewidth{0pt}
 
\tablehead{
\multicolumn{1}{l}{Ion}&
\multicolumn{1}{c}{23 Ori SLV\tablenotemark{b}}&
\multicolumn{1}{c}{$\zeta$ Ori main\tablenotemark{c}}&
\multicolumn{1}{c}{1 Sco main\tablenotemark{d}}&
\multicolumn{1}{c}{$\delta$ Sco main\tablenotemark{e}}&
\multicolumn{1}{c}{$\zeta$ Oph B\tablenotemark{f}}}
\startdata
Li I  & $<9.70$ & \nodata & \nodata &   8.92  &   9.60  \\
Li II & (11.4)  & (11.0)  & (11.8)  & (11.8)  & (11.8)  \\
%$n_e$ & $<0.71$ & \nodata & \nodata &   0.047 &   0.22  \\
$n_e$ & $<0.71$/\nodata & \nodata/\nodata & \nodata/\nodata & 0.047/\nodata & 0.22/\nodata \\
 ~ & \\
C I   &  14.98  &  13.53  &  13.91  &  14.20  &  15.51  \\
C II  &  16.95  & (16.45) &  17.20  &  17.38  &  17.32  \\
%$n_e$ &   0.26  &   0.029 &   0.012 &   0.016 &   0.37  \\
$n_e$ & 0.258/0.214 & 0.029/0.007 & 0.062/0.004 & 0.080/0.020 & 0.373/0.077 \\
 ~ & \\
\enddata
\tablenotetext{a}{Entries for ions are log $N$ (cm$^{-2}$).
References for Li I, C I, Na I, K I may be found in WH; references for S I, Ca I, Ca II, and Fe I are in the appendix to this paper.
Unless otherwise noted below, values for other species are from fits to archival {\it HST}/GHRS spectra.
Values in parentheses are estimated from N(Zn II) and typical depletions. 
Values tabulated here are for the main components in these lines of sight, and thus may be slightly different from the total line of sight values given in the references.
Entries for $n_e$ are cm$^{-3}$, and are calculated via the $\Gamma$/$\alpha_r$ values in Table 3 for $T$ = 100 K and the average interstellar radiation field (except for 1 Sco and $\delta$ Sco). 
First $n_e$ is for radiative recombination only; second $n_e$ includes grain-assisted recombination (see text for details).
}
\tablenotetext{b}{23 Ori:  SLV components at $v_{\odot}$ $\sim$ 20 to 26 km~s$^{-1}$ (Welty et al. 1999b).}
\tablenotetext{c}{$\zeta$ Ori main:  components at $v_{\odot}$ $\sim$ 22 to 27 km~s$^{-1}$.}
\tablenotetext{d}{1 Sco:  components at $v_{\odot}$ $\sim$ $-$9 to $-$5 km~s$^{-1}$.}
\tablenotetext{e}{$\delta$ Sco:  components at $v_{\odot}$ $\sim$ $-$14 to $-$11 km~s$^{-1}$.
C II from Hobbs et al. 1982; Al I from Snow et al. 1986.}
\tablenotetext{f}{$\zeta$ Oph B:  components at $v_{\odot}$ $\sim$ $-$16 to $-$12 km~s$^{-1}$.
Some first ions adapted from Savage et al. 1992; C II from Cardelli et al. 1996; Al I from Shulman et al. 1974; Cr I from Meyer \& Roth 1990.}
\end{deluxetable}

\begin{deluxetable}{lccccccccc}
\tabletypesize{\scriptsize}
\tablecolumns{10}
\tablenum{7}
\tablecaption{Pressure, Density, and Fractional Ionization\tablenotemark{a}\label{tab:pdfi}}
\tablewidth{0pt}
 
\tablehead{
\multicolumn{1}{l}{Star}&
\multicolumn{1}{c}{log[$N$(C I$^0$)]}&
\multicolumn{1}{c}{ }&
\multicolumn{1}{c}{log[$n_{\rm H}T$]\tablenotemark{c}}&
\multicolumn{1}{c}{$T$\tablenotemark{d}}&
\multicolumn{1}{c}{$n_{\rm H}$}&
\multicolumn{1}{c}{log[$N$(Ca I)/$N$(Ca II)]}&
\multicolumn{1}{c}{$n_e$(Ca)\tablenotemark{e}}&
\multicolumn{1}{c}{$n_e$(CNaK)\tablenotemark{f}}&
\multicolumn{1}{c}{$n_e$/$n_{\rm H}$\tablenotemark{g}}\\
\multicolumn{1}{c}{~ ~Component}&
\multicolumn{1}{c}{log[$N$(C I*)]}&
\multicolumn{1}{c}{$f_1$\tablenotemark{b}}&
\multicolumn{1}{c}{}&
\multicolumn{1}{c}{(K)}&
\multicolumn{1}{c}{(cm$^{-3}$)}&
\multicolumn{1}{c}{ }&
\multicolumn{1}{c}{(cm$^{-3}$)}&
\multicolumn{1}{c}{(cm$^{-3}$)}&
\multicolumn{1}{c}{ }\\
\multicolumn{1}{c}{ }&
\multicolumn{1}{c}{log[$N$(C I**)]}&
\multicolumn{1}{c}{$f_2$}&
\multicolumn{1}{c}{ }&
\multicolumn{1}{c}{ }&
\multicolumn{1}{c}{ }&
\multicolumn{1}{c}{ }&
\multicolumn{1}{c}{ }&
\multicolumn{1}{c}{ }&
\multicolumn{1}{c}{ }}
\startdata
o Per\tablenotemark{h} & 15.59 &\nodata& 3.45$\pm$0.1&  50 &  55 & $-$2.08 & 0.342&(0.218)& 0.00396 \\
~ ~ ~main     & 14.88 & 0.158 &     &     &     &     & 0.342&(0.089)&0.00162 \\
              & 14.16 & 0.030 &     &     &     &     &     &     &    \\
              &  \\
$\zeta$ Per\tablenotemark{h} & 15.48 &\nodata& 2.85$\pm$0.1&  55 &  13 & $-$2.16 & 0.304&(0.185)& 0.01420 \\
~ ~ ~main     & 14.25 & 0.054 &     &     &     &     & 0.304&(0.128)&0.00985 \\
              & 14.09 & 0.037 &     &     &     &     &     &     &    \\
              &  \\
\enddata
\tablenotetext{a}{Unless otherwise noted, column density data are from same sources as in Table 6.}
\tablenotetext{b}{Fractions of C I in first and second excited states.}
\tablenotetext{c}{Estimated from graphs in Jenkins \& Shaya 1979.}
\tablenotetext{d}{From H$_2$ rotational excitation (Savage et al. 1977; Rachford et al. 2002); values in parentheses are assumed.}
\tablenotetext{e}{Corrected for $T$ $<$ 100 K; WJ1 radiation field assumed (multiplied by 5 for Scorpius sightlines).  First line for radiative recombination only; second line includes grain-assisted recombination.}
\tablenotetext{f}{Average value for C, Na, and K; corrected for $T$ $<$ 100 K; WJ1 radiation field assumed (multiplied by 5 for Scorpius sightlines).  First line for radiative recombination only; second line includes grain-assisted recombination; values in parentheses are for entire line of sight.}
\tablenotetext{g}{Using $n_e$ for C, Na, K.  First line for radiative recombination only; second line includes grain-assisted recombination.}
\tablenotetext{h}{C I fine-structure analysis from 
Wannier et al. 1999 (o Per, $\zeta$ Per);
Zsarg\'{o} 2000 ($\lambda$ Ori, $\sigma$ Sco);
Lambert et al. 1994 ($\zeta$ Oph);
Sonnentrucker et al. 2002 (HD 192639) and in prep. (HD 185418); and
Jenkins \& Tripp 2001 ($\lambda$ Cep).}
\end{deluxetable}

\begin{deluxetable}{lrrrrrc}
\tabletypesize{\scriptsize}
\tablecolumns{7}
\tablenum{8}
\tablecaption{Fe~I $\lambda$3719 Component Structures\label{tab:fe1}}
\tablewidth{0pt}
 
\tablehead{
\multicolumn{1}{l}{Star}&
\multicolumn{1}{c}{S/N}&
\multicolumn{1}{c}{W$_{\lambda}$\tablenotemark{a}}&
\multicolumn{1}{c}{Comp}&
\multicolumn{1}{c}{$v_\odot$}&
\multicolumn{1}{c}{$N$}&
\multicolumn{1}{c}{$b$\tablenotemark{b}}\\
\multicolumn{1}{c}{ }&
\multicolumn{1}{c}{ }&
\multicolumn{1}{c}{(m\AA)}&
\multicolumn{1}{c}{ }&
\multicolumn{1}{c}{(km~s$^{-1}$)}&
\multicolumn{1}{c}{(10$^{10}$ cm$^{-2}$)}&
\multicolumn{1}{c}{(km~s$^{-1}$)}}
\startdata
$\zeta$ Ori    & 245&        $<$0.1& & \nodata& $<$2.0&\nodata \\
               &    &              & &        &       &        \\
HD 72127 A     & 105&        $<$0.3& & \nodata& $<$6.0&\nodata \\
               &    &              & &        &       &        \\
$\rho$ Leo     & 105&        $<$0.3& & \nodata& $<$6.0&\nodata \\
               &    &              & &        &       &        \\
$\theta$ Car   & 265&        $<$0.1& & \nodata& $<$2.0&\nodata \\
               &    &              & &        &       &        \\
\enddata
\tablenotetext{a}{Uncertainties and limits are 2$\sigma$.
Chaffee 1974 found 3$\pm$1 m\AA\ for $\zeta$ Per, 5$\pm$2 m\AA\ for o Per, and $<$1 m\AA\ for $\xi$ Per.
Snow et al. 1986 found 1.0$\pm$0.8 m\AA\ for $\sigma$ Sco, $<$1 m\AA\ for $\delta$ Sco, and 13$\pm$2 m\AA\ for $\rho$ Oph.
Shulman et al. 1974 found 1.9 m\AA\ for $\zeta$ Oph.}
\tablenotetext{b}{Values of $b$ in square brackets were fixed to facilitate convergence of the fits, but are well-determined.  Values of $b$ in parenthese were also fixed, but are less well-determined.}
\tablenotetext{c}{Equivalent width and column densities from fit to profile shown by Barlow et al. 1995, who gave $b$-values and velocities.}
\end{deluxetable}

\begin{deluxetable}{rlcrrrrrrrr}
\tabletypesize{\scriptsize}
\tablecolumns{11} 
\tablenum{9}
\tablecaption{Column Densities for Ca~I, Ca~II, Fe~I, and S~I\tablenotemark{a}\label{tab:coldens}}
\tablewidth{0pt} 
 
\tablehead{
\multicolumn{1}{c}{HD}&
\multicolumn{1}{c}{Name}&
\multicolumn{1}{c}{WH\tablenotemark{b}}&
\multicolumn{1}{c}{Ca I\tablenotemark{c}}&
\multicolumn{1}{c}{Ref\tablenotemark{c}}&
\multicolumn{1}{c}{Ca II\tablenotemark{d}}&
\multicolumn{1}{c}{Ref\tablenotemark{d}}&
\multicolumn{1}{c}{Fe I\tablenotemark{e}}&
\multicolumn{1}{c}{Ref\tablenotemark{e}}&
\multicolumn{1}{c}{S I\tablenotemark{f}}&
\multicolumn{1}{c}{Ref\tablenotemark{f}}}
\startdata
  2905 & $\kappa$ Cas  &*&   9.92 & 1&  12.70 & 1& \nodata&  &  13.38 & 1\\
  5394 & $\gamma$ Cas  &*& $<$9.86& 2&  11.16 & 2& \nodata&  & \nodata&  \\
 20041 &               & &  10.34 & 3& \nodata&  &$<$12.56& 1& \nodata&  \\
 21291 &               &*&  10.33 & 1& \nodata&  &  12.25 & 2& \nodata&  \\
 21389 &               & &  10.38 & 4& \nodata&  &  12.25 & 2& \nodata&  \\
 21483 &               & &  10.24 & 4& \nodata&  &  12.02 & 2& \nodata&  \\
 23180 & o Per         &*&   9.87 & 1&  12.09 & 2&  11.95 & 1&  13.74 & 1\\
281159 & BD+31 643     & &  10.07 & 3& \nodata&  & \nodata&  & \nodata&  \\
\enddata

\tablenotetext{a}{Entries are log($N$) (cm$^{-2}$), and refer to the total value for the entire sightline.  A colon indicates values for which the 1$\sigma$ uncertainty is greater than 25\%.  Upper limits are 3$\sigma$.}
\tablenotetext{b}{Sightlines included in the appendix to Welty \& Hobbs 2001 are denoted by asterisks.}
\tablenotetext{c}{Ca I:  1 = this paper; 2 = Allen 1994; 3 = Thorburn, priv. comm.; 4 = Welty unpubl.; 5 = White 1973; 6 = Penprase 1993; 7 = Welty et al. 1999b; 8 = Lambert \& Danks 1986; 9 = Knauth et al. 2001; 10 = Gredel et al. 1993; 11 = Cardelli et al. 1990; 12 = Crawford 2002; 13 = Cardelli \& Wallerstein 1986; 14 = Haffner \& Meyer 1995; 15 = Federman \& Hobbs 1983; 16 = Hanson et al. 1992; 17 = Hobbs 1971; 18 = Federman et al. 1997.}
\tablenotetext{d}{Ca II:  1 = Welty, unpubl.; 2 = Welty et al. 1996; 3 = Hobbs 1974, 1978: 4 = Welty et al. 1999b; 5 = Price et al. 2001: 6 = Knauth et al. 2001; 7 = Hobbs et al. 1991; 8 = Crawford 2002; 9 = Penprase 1993; 10 = Crawford 1995; 11 = Sahu et al. 1998.}
\tablenotetext{e}{Fe I:  1 = Thorburn, priv. comm.; 2 = Joseph et al. 1989; 3 = Chaffee 1974; 4 = Welty et al. 1999b; 5 = this paper; 6 = Cardelli et al. 1990; 7 = Welty, unpubl.; 8 = Cardelli \& Wallerstein 1986; 9 = Haffner \& Meyer 1995; 10 = Barlow et al. 1995; 11 = Snow et al. 1996.}
\tablenotetext{f}{S I:  1 = Welty et al. 1991; 2 = Welty, unpubl.; 3 = Welty et al. 1999b; 4 = Allen et al. 1990; 5 = Zsarg\'{o} et al. 1997; 6 = Federman \& Cardelli 1995; 7 = Snow et al. 1996; 8 = Hanson et al. 1992; 9 = Sonnentrucker et al. 2002 and in prep.}
\end{deluxetable}

\end{document}